\documentclass[preprint,showkeys,superscriptaddress,nofootinbib%
]{revtex4-1}

\usepackage{amsmath,amssymb,epsfig,color}
\usepackage{booktabs}
\usepackage{bm}

\allowdisplaybreaks[4]
\newcommand{\lsim}{%
\raise0.3ex\hbox{$\;<$\kern-0.75em\raise-1.1ex\hbox{$\sim\;$}}}
\newcommand{\gsim}{%
\raise0.3ex\hbox{$\;>$\kern-0.75em\raise-1.1ex\hbox{$\sim\;$}}}

\newcommand{\eq}[1]{(\ref{#1})}
\newcommand{\Eq}[1]{Eq.~(\ref{#1})}
\newcommand{\eqs}[2]{(\ref{#1})--(\ref{#2})}
\newcommand{\Eqs}[2]{Eqs.~(\ref{#1})--(\ref{#2})}

\newcommand{\eg}{{\it e}.{\it g}.}
\newcommand{\abs}[1]{\left| #1 \right|}

\newcommand{\wt}[1]{{\widetilde #1}}
\newcommand{\VEV}[1]{\left<#1\right>}
\newcommand{\order}[1]{{\cal O}\!\left(#1\right)}
\newcommand{\fun}[1]{\!\left(#1\right)} 
\newcommand{\ie}{{\it i}.{\it e}.}
\newcommand{\units}[1]{\,\mathrm{[#1]}} 
\newcommand{\nunits}[1]{\,\mathrm{#1}}  

\newcommand{\wb}[1]{%
\vbox{\ialign{##\crcr\hskip 1.0pt\hrulefill\hskip 0.3pt%
\crcr\noalign{\kern-1pt\vskip0.07cm\nointerlineskip}%
$\hfil\displaystyle{#1}\hfil$\crcr}}}

\newcommand{\MS}{M_S} 
\newcommand{\ME}{M_{E}} 

\newcommand{\Tr}[1]{\mathrm{Tr}\!\left(#1\right)}
\renewcommand{\Tr}[1]{\mathrm{Tr}\,{#1}}
\renewcommand{\Tr}[1]{\mathrm{Tr}\big(#1\big)}

\newcommand{\suppress}[1]{\varepsilon_{#1}} 
\newcommand{\suppressionD}[1]{\suppress{#1}} 

\newcommand{\lambdaad}{\lambda_A}

\renewcommand{\paragraph}[1]{
\medskip\medskip
\noindent\underline{\hbox{\bf #1}}
}

\begin{document}

\title{%
Next-to-minimal $R$-symmetric model:\\
Dirac gaugino, Higgs mass and invisible width
}

\author{Hiroaki {\sc Nakano}}
\email{nakano@muse.sc.niigata-u.ac.jp}
\affiliation{Department of Physics, 
Niigata University,~Niigata, 950-2181, Japan}

\author{Masaki {\sc Yoshikawa}}
\email{yosikawa@muse.sc.niigata-u.ac.jp}
\affiliation{Graduate School of Science and Technology, 
Niigata University,~Niigata, 950-2181, Japan\\[20pt]
\medskip
}

\begin{abstract}%
We study a singlet extension of
the minimal ${U(1)}_R$ symmetric model,
which 
shares nice properties of Dirac gauginos
and $R$-symmetric Higgs sector.
At the same time,
a superpotential coupling of $R$-charged singlet
to the Higgs doublets can give a substantial contribution
to the Higgs boson mass.
We show that 
the 125 GeV Higgs boson is consistent with
perturbative unification,
even if the SUSY scale is as low as 1 TeV
and if the $D$-term Higgs potential is suppressed 
as is often the case in Dirac gauginos.
The model also contains a light scalar and fermion,
pseudo-moduli and pseudo-Goldstino:
The former gets a mass mainly from SUSY breaking soft terms,
in addition to a small explicit $R$-symmetry breaking
for the latter.
We examine how the Higgs mass and width
are affected by these light degrees of freedom.
Specifically we find that
depending on parameters of $R$-charged Higgses,
the pseudo-moduli lighter than a half of the SM-Higgs boson mass
is still allowed by the constraints from 
invisible decays of the $Z$ and Higgs bosons.
We also find that
such a light scalar can reduce the Higgs boson mass,
at most by a few percents.
\end{abstract}

\maketitle


\section{Introduction}
\label{sec:introduction}

Low--energy supersymmetry (SUSY) is a fascinating idea
since it was proposed as a solution to the hierarchy problem
and further supported by a sign of gauge coupling unification.
Such idea is now challenged, however,
since no signal for SUSY particles has been 
reported yet
at the Large Hadron Collider (LHC).
The discovery of the Standard Model (SM)-like Higgs boson
at $125\nunits{GeV}$ also indicates that
SUSY would be heavy \cite{Giudice:2011cg}, if exists,
since the identification of the discovered boson
with the lightest Higgs boson in the Minimal SUSY SM (MSSM) 
requires large radiative corrections from the top squarks.

An $R$-symmetric realization of low-energy SUSY 
provides an alternative to the MSSM \cite{Hall:1990hq,Randall:1992cq}.
In particular,
models with Dirac gaugino have recently attracted much attention
in several respects:
``supersoftness'' of scalar masses \cite{Fox:2002bu}, 
``supersafeness'' of squark production 
\cite{Heikinheimo:2011fk,Kribs:2012gx},
as well as 
improved properties in flavor sector \cite{Kribs:2007ac}:
The adjoint chiral multiplets,
including Dirac partners of the MSSM gauginos,
also contain adjoint scalars, 
$\mathcal{N}=2$ partners of the gauginos,
which have a suitable coupling to cancel UV divergences
of scalar masses.
This can improve naturalness property of the Higgs potential.

In the minimal setup of $U(1)_R$-symmetric model \cite{Kribs:2007ac},
we introduce an $R$-partner to each gaugino and Higgsino.
An $R$-partner of the MSSM gaugino $\lambda_{a=3,2,1}$,
is contained in an adjoint chiral multiplet,
$\mathcal{A}_3=O$, $\mathcal{A}_2=T$ and $\mathcal{A}_1=S$,
while an $R$-partner of the MSSM Higgs doublet, $H_{i=u,d}$,
is denoted by $R_{i=u,d}$.
The minimal $R$-symmetric model is known to have advantages
of suppressing flavor and CP violations 
\cite{Hisano:2006mv,Kribs:2007ac,Fok:2010vk,Dudas:2013gga};
Phenomenological study \cite{Choi:2010an}
and electroweak (EW) baryogenesis \cite{Kumar:2011np,Fok:2012fb}
was also discussed.
See Refs.~\cite{Frugiuele:2012pe,Morita:2012kh,Chakraborty:2013gea} 
for $R$-symmetric models with different $R$-charge assignments,
and Ref.~\cite{Benakli:2014cia}
for a minimal SUSY model of Dirac gaugino
with and without $R$-symmetry.

An important step was made 
in Refs.~\cite{Bertuzzo:2014bwa,Diessner:2014ksa} showing that
the $125\nunits{GeV}$ Higgs mass can be accommodated
if the $SU(2)\times U(1)$ adjoints $\mathcal{A}_{1,2}$ have
sufficiently large couplings to the Higgs pairs $R_iH_i$:
Such adjoint Yukawa couplings can also relax 
the suppression of $SU(2)\times U(1)$ $D$-term potential,
which is known as a drawback of supersoft SUSY breaking models.
Notice, however, that
although marginally allowed \cite{Diessner:2015yna},
such large adjoint couplings are generally in tension with 
EW precision measurements,
as in the muless SUSY SM \cite{Nelson:2002ca}.
Perturbativity constraints should also be taken into account.

In the present paper,
we consider a singlet extension of 
the minimal $R$-symmetric SUSY SM, 
which may be called 
the next--to--minimal $R$-symmetric model for short;
It is actually a combination of Dirac gaugino
and an $R$-symmetric Higgs sector
proposed in Ref.~\cite{Izawa:2011hi}.
Unlike the conventional singlet extensions of the MSSM,
the present model is not intended to explain
the origin of Higgsino mass terms.
Instead, we would like to examine 
how the lightest Higgs mass of $125\nunits{GeV}$
can be reproduced in our $R$-symmetric setup.
Actually we show that
$125\nunits{GeV}$ Higgs mass 
can well be accommodated
by a singlet Yukawa coupling to the MSSM Higgs pair,
consistently with perturbative unification.
We also examine to what extent
the adjoint Yukawa couplings can affect the results.
As is expected,
our solution requires small $\tan\beta$
so that the singlet coupling can give a substantial contribution
to the lightest Higgs mass;
it also favors small values for the adjoint Yukawa couplings.
Therefore our setup provides 
an alternative to the known solution in the minimal $R$-symmetric model.

In the present work, 
we assume tha explicit $R$ symmetry breaking
is very small, as was discussed in Ref.~\cite{Morita:2012kh};
Through coupling to supergravity,
the theory contains a small explicit breaking of $R$-symmetry,
characterized by the gravitino mass,
of the order of a few GeV, for instance.
Throughout the present paper,
we will neglect such explicit $R$-breaking effects,
including anomaly-mediated contribution to Majorana gaugino masses.
Cosmological issues will be discussed elsewhere.

The most characteristic feature of the present model 
is the presence of light degrees of freedom,
pseudo-moduli $\phi$ and pseudo-Goldstino $\psi$.
In Ref.~\cite{Bertolini:2011tw},
properties of such light degrees of freedom
were studied in some details;
A generic prediction there is that
the light scalar $\phi$ couples to the Higgs boson so strongly that
the Higgs decays almost invisibly to its pair.
We revisit this issue here,
and show that
there is a region of parameter space in which
a light scalar mode still exists
and, unlike the previous expectation,
the Higgs invisible width is within the present bound.
We explain why and when the coupling of the light scalars
to the Higgs boson becomes weak
thanks to its pseudo-moduli nature.



It is worth mentioning that
the present model has a similarity
with the conventional singlet extensions of the MSSM
--- the NMSSM
\cite{Ellwanger:2009dp,Badziak:2013bda,Ellwanger:2012ke,%
Kobayashi:2012ee,Goodsell:2014pla},
the nMSSM \cite{%
Panagiotakopoulos:1999ah,Panagiotakopoulos:2000wp,Ishikawa:2014owa}
or the PQ-NMSSM \cite{Jeong:2011jk,Jeong:2012ma,Choi:2013lda}:
the SM-like Higgs boson mass can be enhanced
at small $\tan\beta$,
but such enhancement is limited by the triviality bound
of the singlet coupling(s).
There are some important differences, however.
First of all,
in the present setup,
the SM-like Higgs boson should be the lightest mass eigenstate
in the $R$-neutral scalars,
since the approximate $R$-symmetry forbids the mixing
to the $R$-charged scalars.
The $R$-symmetry also forbids $A$-terms 
that could be used for enhancing the Higgs mass.
Moreover,
in the conventional models,
the Higgsino mass term is forbidden 
by imposing some symmetries
while our model contains $R$-invariant Higgsino mass parameters
from the start.
These mass terms, on a theoretical side,
trigger spontaneous SUSY breaking in the Higgs sector
\cite{Izawa:2011hi};
At the same time, the mass terms 
can suppress (unwanted) mixings between
the singlet and doublet states in the $R$-charged sector.
We thus expect that
the resulting phenomenology can be quite different.


The present paper is organized as follows.
In \S\ref{sec:setup},
we introduce out setup for a singlet extension of the minimal
$R$-symmetric SUSY SM and describe its characteristic features,
the existence of light degrees of freedom 
related to a pseudo-moduli direction.
We also give a brief review of aspects of Dirac gaugino
in supersoft SUSY breaking,
including the suppression of $D$-term quartic scalar potential.
In \S~\ref{sec:triviality} 
we discuss how the $125\nunits{GeV}$ Higgs boson can be compatible
with perturbative unification.
We use two-loop renormalization group equations (RGE's)
but mainly consider the case
in which the SUSY particle masses
can be treated 
by a single mass threshold $\MS$.
We calculate the lightest Higgs mass
in the RG approach:
At the scale $\MS$, we match the $R$-symmetric model
to the minimal SM
by taking the decoupling limit of the heavier Higgs states.
We also take into account various effects
including adjoint Yukawa interactions and 
a possibility of a light pseudo-moduli below the matching scale.

The constraints from 
invisible decay of the $Z$ and SM-like Higgs bosons
are examined in \S~\ref{sec:decay}.
The constraint from the $Z$ decay to a pseudo-Goldstino pair
can be satisfied if 
the $R$-symmetric Higgsino mass parameters are not too small.
We also examine the invisible Higgs decay to a pseudo-moduli pair,
and find that
there is a region of parameter space
corresponding to a pseudo-moduli that 
is lighter than a half of the SM-like Higgs boson mass,
and such a light pseudo-moduli has very suppressed coupling
to the Higgs boson(s).
The final section is devoted to conclusion and discussion.
Necessary tools for the RG analysis
and some detailed discussion about 
gauge coupling unification and 
Dirac gaugino mass threshold
are given in Appendices.

\section{Dirac Gaugino and 
Next-to-minimal $R$-symmetric Higgs Sector}
\label{sec:setup}

We consider the theory invariant under $U(1)_R$
under which all the SM fields are neutral.
The $R$-symmetric superpotential proposed in Ref.~\cite{Izawa:2011hi}
is given by $W = W_H + W_Y$, where\,\footnote{%
The $R$-charged fields $R_0$, $R_u$ and $R_d$
were denoted in Ref.~\cite{Izawa:2011hi}
by $X_0$, $X_1$ and $X_2$, respectively.
Accordingly the Higgsino mass parameters
$\mu_u$ and $\mu_d$ here correspond to $\mu_1$ and $-\mu_2$ there;
we have flipped the sign of $\mu_d$ term
so that
all the neutral components have the same sign.
Note also that the parameters, $\lambda$, $\mu_{u,d}$, $f$, 
can be made real and positive
by field redefinition of $R_{0,u,d}$ and $H_{u,d}$.
} 
\begin{align} 
  W_H
  &= R_0\left(f+\lambda H_uH_d\right)
   + \mu _uR_uH_u-\mu_dR_dH_d
   \ ,
 \label{eq:W:Higgs} 
\\
 W_Y
 &= y^{ij}_u Q_iU^c_jH_u + y^{ij}_d Q_iD^c_jH_d + y^{ij}_e E^c_iL_jH_d
 \ .
 \label{eq:W:Yukawa} 
\end{align}
Here $H_{u,d}$ are the MSSM Higgs doublets,
$R_{u,d}$ are their $R$-partners of $R$ charge $+2$, and
$\mu_{u,d}$ are $R$-invariant Higgsino mass parameters
which we assume to be of the order of the weak scale.
A dimension--two parameter $f$ of the order of the weak scale 
can act as additional source of (super)symmetry breaking,
as was further studied in Refs.~\cite{Bertolini:2011tw,Morita:2012kh}.
A possible origin of these mass parameters
was also discussed in Ref.~\cite{Izawa:2011hi}.
In addition, we introduce a gauge singlet $R_0$ of $R$-charge 2,
which has an nMSSM-like coupling $\lambda$
to the MSSM Higgs doublets $H_{u,d}$.
We will refer to this as \textit{singlet Yukawa coupling}.
It is the purpose of the present paper
to discuss the implications of this singlet Yukawa coupling
in the Higgs mass and decay width.

The $U(1)_R$ symmetry forbids the usual Majorana mass terms
for the MSSM gauginos.
To allow Dirac gaugino mass terms
we introduce chiral supermultiplets $\mathcal{A}_a$
belonging to the adjoint representation of
the SM gauge group $SU(3)_c \times SU(2)_L \times U(1)_Y$,
where $a=3,2,1$ is a gauge index;
Explicitly,
$\mathcal{A}_3=O$ is an $SU(3)$-octet, 
$\mathcal{A}_2=T$ an $SU(2)$-triplet, $\mathcal{A}_1=S$ a singlet.
We denote their fermion and scalar components
by $\chi_a$ and $A_a=\left(\sigma_a+i\pi_a\right)/\sqrt{2}$, respectively.
Matter contents and $R$-charge assignment are 
summarized in the Table.~\ref{tab:contents}.

\begin{table}[tbp]
\begin{center}
\begin{tabular}{c|cc|ccc|cc|cc}
 & \makebox[9mm]{ $H_u$ }
 & \makebox[9mm]{ $H_d$ }
 & \makebox[9mm]{ $R_u$ }
 & \makebox[8mm]{ $R_d$ }
 & \makebox[9mm]{ $R_0$ }
 & \makebox[6mm]{ $S$ }
 & \makebox[6mm]{ $T$ }
 & \makebox[10mm]{ $E^c_{4,5}$   }
 & \makebox[10mm]{ $\wb{E}^c_{4,5}$   } \\ \hline \hline
 $SU(2)_L$ 
 & $\mathbf 2$ & $\mathbf 2$ 
 & $\mathbf 2$ & $\mathbf 2$ 
 & $\mathbf 1$ & $\mathbf 1$ 
 & $\mathbf 3$ 
 & $\mathbf 1$ & $\mathbf 1$ \\
$U(1)_Y$ 
 & $+1/2$ & $-1/2$ 
 & $-1/2$ & $+1/2$ &  $0$ & $0$ &  $0$ & $+1$ & $-1$ \\ \hline
$U(1)_R$
 &  $0$ & $0$ 
 & $+2$ & $+2$ &   $+2$ &    $0$  &    $0$ 
 & $1+r$ & $1-r$ \\
\end{tabular}
\end{center}
\caption{
The charge assignments of the Higgs and extra fields,
other than the $SU(3)$-octet $\mathcal{A}_3=O$.
The MSSM Higgs doublets are $R$-neutral while 
their $R$-partners as well as the singlet $R_0$ have $R$-charge $+2$.
The adjoint chiral multiplets $\mathcal{A}_a$ are $R$-neutral
so that their fermionic components have $R$-charge $-1$.
All the quark and lepton superfields have $R$-charge $+1$.
We also introduce two pairs of vector-like ``leptons'',
as is discussed in \S{\ref{subsec:unification}}.
}
\label{tab:contents}
\end{table}

In the presence of these adjoint chiral multiplets,
we can add the following superpotential\,\footnote{
Following the existing literature, we put a normalization factor
$k_{2}=2$ for $SU(2)$-adjoint Yukawa terms while $k_{1}=1$ for $U(1)$.
We also put sign factors $\eta_u=+1$ and $\eta_d=-1$
so that the neutral components have plus signs.
}
\begin{align} 
 W_A
 &=\sum_{a=2,1}\sum_{i=u,d}
   \eta_i k_a\lambda ^i_a R_i\mathcal{A}_a H_i
\nonumber\\
 &= 2\lambda ^u_T R_u T H_u - 2\lambda ^d_T R_d T H_d 
   + \lambda ^u_S R_u S H_u - \lambda ^d_S R_d S H_d 
   \ ,
\label{eq:W:adjoint} 
\end{align}
which we refer to as \textit{adjoint Yukawa terms}
(although $\mathcal{A}_1=S$ is actually a singlet).
In Refs.~\cite{Bertuzzo:2014bwa,Diessner:2014ksa},
these terms play a central role
in reproducing the lightest Higgs mass of $125\nunits{GeV}$.
In the present model, however,
it is not these adjoint Yukawa terms but 
the singlet Yukawa term 
in \Eq{eq:W:Higgs} 
that is important for the Higgs mass.

The superpotential $W=W_H+W_Y+W_A$
is not completely general one
that is allowed by the $R$-symmetry.
Our assumption here is that
there is no superpotential term
that is quadratic in adjoint chiral multiplets $\mathcal{A}_a$,
such as $R_0S^2$ and $R_0T^2$.
We also assume that
the mixing mass term of $S$ and $R_0$ 
does not arise after GUT symmetry breaking.
To justify these assumptions would require 
a concrete embedding of the model into grand unified theories (GUT's):
Here we just note that an interesting possibility is provided
by an orbifold-type model \cite{Kojima:2011ad},
which does contain light chiral adjoints after GUT symmetry breaking.


\subsection{Supersoft SUSY Breaking and Suppression of $D$-term} 
\label{subsec:supersoft}

Dirac gaugino mass terms are generated through
so-called supersoft operator~\cite{Polchinski:1982an,Fox:2002bu}
\begin{align}
 \mathcal{L}_{\mathrm{SS}}
 = \sum_{a=3,2,1}\int d^2\theta 
   \frac{\sqrt{2}\,g_a}{\Lambda _D}\,
   \mathcal{W}^\alpha_X \mathcal{W}^a_\alpha \mathcal{A}_a
 + \mathrm{H.c.} \ ,
 \label{eq:DG:superpot}
\end{align}
where 
$\mathcal{W}_X$ is the gauge field strength of a hidden sector $U(1)_X$ 
whose nonvanishing $D$-term\,\footnote{
See Refs.~\cite{Itoyama:2011zi,Itoyama:2012fk}
for a mechanism of $D$-term dynamical SUSY,
and Refs.~\cite{Alves:2015kia,Alves:2015bba}
for a mechanism for generating the supersoft operator
through a Wess--Zumino--Witten term.
Note that a constant shift in adjoint scalars
is not allowed by the supersoft operator \eq{eq:DG:superpot}.
} 
is a source of SUSY breaking
and $\Lambda_D$ is a cutoff scale 
at which the above operator is generated.
A vacuum expectation value (VEV) of $U(1)_X$ $D$-term, 
$\VEV{D_X}$, gives
\begin{align} 
 \mathcal{L}_{\mathrm{SS}}
 \longrightarrow{} 
 - \Bigl\{
	 m_{D_a}\lambda_a \chi_a + \mathrm{H.c.}
   \Bigr\}
	 - 2m_{D_a} D_a \sigma_a 
  \ ,
 \label{eq:DG:lag}	 
\end{align}
where $m_{D_a}=g_a\VEV{D_X}/\Lambda_D$ is a Dirac mass of 
gaugino $\lambda_a$ and adjoint fermion $\chi_a$
(at the scale $\Lambda_D$);
At the same time, the second term gives rise to 
the mass term of the real part $\sigma_a$ of the adjoint scalar and 
the trilinear scalar interactions of $\sigma_a$ to other scalars.
The latter has important consequences,
supersoftness and $D$-term cancellation,
which we review shortly.

If the $U(1)_X$ $D$-term is the only source of SUSY breaking,
a soft scalar mass in the visible sector is UV finite.
Such supersoft contribution is given by \cite{Fox:2002bu}
\begin{align}
\delta m^2\fun{\varphi_i}
 &= \sum_{a=3,2,1}
    \frac{4C_2^a\fun{\varphi_i}g^2_a}{16\pi^2}\,m_{D_a}^2
    \log\frac{m_{\sigma_a}^2}{m_{D_a}^2}
    \ ,
\label{eq:supersoft:scalarmass}
\end{align}
where $C_2$ is a quadratic Casimir for a scalar $\varphi_i$ and 
$m^2_{\sigma_a}$ is the mass squared of 
the adjoint scalar $\sigma_a$.
Note that its pseudo-scalar partner $\pi_a$ 
does not receive a mass from the supersoft operator \eq{eq:DG:superpot}.
Although the $U(1)_R$ symmetry forbids 
supersymmetric mass term of $\mathcal{A}_a$,
its scalar components 
$A_a=\left(\sigma_a+i\pi_a\right)/\sqrt{2}$ can get a mass 
from holomorphic and non-holomorphic soft mass terms
\cite{Carpenter:2010as,Csaki:2013fla,Alves:2015kia,Alves:2015bba}.
In addition,
$SU(3)\times SU(2)$ adjoints receive 
finite loop corrections \eq{eq:supersoft:scalarmass} to their masses.
We denote the resultant masses of adjoint scalars
by $m_\sigma^2=4m_D^2+\delta m_\sigma^2$ and $m_\pi^2$, respectively.
We will assume that
these adjoint scalars have large enough masses 
to prevent them from developing a nonzero VEV
especially in the presence of the adjoint Yukawa terms
\eq{eq:W:adjoint}. 

Another characteristic feature of Dirac gaugino models
is the suppression of the $D$-term scalar potential \cite{Fox:2002bu}.
Here let us recall it in the absence of adjoint Yukawa terms.
Then the relevant terms are
\begin{align} 
 V_D
  =  \sum_{a} \frac{1}{2}
     \left[ 
       2m_{D_a} \sigma _a - \sum_i \varphi^*_i g_a T^a \varphi_i
     \right]^2
   + \sum_a\!\left[
        \frac{1}{2}\,\delta m_{\sigma_a}^2 \left(\sigma_a\right)^2
      + \frac{1}{2}\,m_{\pi_a}^2   \left(\pi_a\right)^2
     \right]
     \ ,
\end{align}
where $\varphi_i$ is a generic scalar field.
If the adjoint scalar $\sigma_a$ can be regarded as heavy enough
to be integrated out,
the effective $D$-term potential is given by
\begin{align} 
 V_D^{\mathrm{eff}}
\ =\ \sum_{a}\suppressionD{D_a}
     \frac{g_a^2}{2}
     \left[\sum_i \varphi^*_i T^a \varphi_i\right]^2
     \ , \qquad
\suppressionD{D_a}
\ \equiv\ \frac{\delta m^2_{\sigma_a}}{4m^2_{D_a}+\delta m^2_{\sigma_a}}
     \ ,
 \label{eq:Dterm:pot}
\end{align}
where $\suppressionD{D_a}$ is a suppression factor
between $0$ and $1$.
Such suppression of the quartic $D$-terms 
is not welcome in low-scale supersymmetry\,\footnote{
See however Refs.~\cite{Unwin:2012fj,Fox:2014moa} for an interesting application.
}
and can be avoided by assuming
large additional contribution $\delta m^2_{\sigma_a}$
to the adjoint scalar masses.
[Of course, this can be done 
at the cost of losing the supersoftness 
\cite{Arvanitaki:2013yja,Csaki:2013fla}.]
Alternatively, 
the adjoint Yukawa terms can relax 
the $D$-term suppression \cite{Bertuzzo:2014bwa,Diessner:2014ksa},
depending on the size and sign of a combination 
$\lambda_{ai}\mu_i/g_a m_{D_a}$.

Later we will consider the common value of 
$\suppressionD{D_1}=\suppressionD{D_2}
\left(=\suppressionD{D}\right)$
and treat it as a free parameter.

\subsection{Pseudo moduli and pseudo Goldstino} 
\label{subsec:pmoduli}

The scalar potential in the $R$-symmetric Higgs sector takes the form
\begin{align} 
  V\ =\ V_F+V_A+V_D^{\mathrm{eff}}+V_{\mathrm{soft}} \ .
\label{eq:V}
\end{align}
Here $V_F$ and $V_A$  correspond, respectively,
to the superpotentials \eq{eq:W:Higgs} 
and \eq{eq:W:adjoint}, 
\begin{align} 
 V_F 
\ = \sum_{I=0,u,d}\abs{F_{Ri}}^2 + \sum_{i=u,d}\abs{F_{Hi}}^2
    \ , \qquad
 V_A
\ =\ \sum_{a=1,2}\abs{F_{A_a}}^2
     \ ,
\nonumber
\end{align}
which read,
explicitly for the neutral components,
\begin{align} 
 V_F 
 =&  \abs{f-\lambda H^0_u H^0_d}^2
   + \abs{\mu_d H^0_d}^2
   + \abs{\mu_u H^0_u}^2
   + \abs{\mu_u R^0_u-\lambda  R_0 H^0_d}^2
   + \abs{\mu_d R^0_d-\lambda R_0 H^0_u}^2
    \,,
 \label{eq:V:RH}\\
 V_A
 =& \sum_{a=1,2}
     \abs{
       \lambda_{ai} R_i^0 H_i^0}^2
\ =\ \left|\lambda ^u_S H^0_u R^0_u + \lambda ^d_S H^0_d R^0_d\right|^2
    +\left|\lambda ^u_T H^0_u R^0_u + \lambda ^d_T H^0_d R^0_d\right|^2
     \ .
 \label{eq:V:adjoint}
\end{align}
The last term $V_{\mathrm{soft}}$ stands for soft scalar masses,
\begin{align} 
V_{\mathrm{soft}}
 &=  \sum_{\varphi_i}m^2_{\varphi_i}\abs{\varphi_i}^2
     \ ,
 \label{eq:V:soft}
\end{align}
which can be induced from any $R$-invariant mediation of SUSY breaking
as in Refs.~\cite{Nelson:2002ca,Blechman:2009if,Kribs:2010md}.
Notice that
in \Eq{eq:V:RH},
we have put adjoint scalar VEV's to zero
by assuming large adjoint scalar masses.



As we mentioned in \S\ref{sec:introduction},
the model contains light degrees of freedom,
pseudo-moduli $\phi$ and pseudo-Goldstino $\psi$.
One way to see their existence is to realize that
the superpotential \eq{eq:W:Higgs} alone
defines a kind of O'Raifeartaigh model 
and that 
the corresponding  scalar potential $V_F$ has a flat direction,
called pseudo moduli direction in the context of 
spontaneous supersymmetry breaking~\cite{Intriligator:2007cp}.
In the present case,
it lies in the space of $R$-charged Higgs fields $R_{0,u,d}$
and can be parametrized by polar angles as
\begin{align} 
 \left(R_0,R^0_u,R^0_d\right)
\ \sim\ 
        \left(
        \cos\theta ,\  \sin\theta \cos\varphi , \ \sin\theta
        \sin\varphi
        \right)
 \ ,
 \label{eq:moduli:state}
\end{align}
where, 
with $\VEV{H_u^0}=v_u=v\sin\beta$ and $\VEV{H_d^0}=v_d=v\cos\beta$,
\begin{align}
 \tan\theta
 &= \frac{\lambda v}{\mu}
  \equiv
  \lambda v\sqrt{
    \frac{\cos^2\beta}{\mu^2_u} + \frac{\sin^2\beta }{\mu^2_d}
    }
 \ ,\qquad
 \tan\varphi
 = \frac{\mu_u}{\mu_d}\tan\beta
  \ . 
 \label{eq:angle:def}
\end{align}
\Eq{eq:moduli:state} is not a true flat direction, 
and is lifted by other terms in the scalar potential \eq{eq:V}.
To discuss its impact on Higgs phenomenology
is another purpose of the present paper.



Let us first discuss the pseudo-Goldstino.
The present model contains nine components of neutralinos:
two gauginos $\lambda_{a=1,2}^0$
and three $R$-charged Higgsinos $\wt{R}_{0,u,d}^0$
of $R$-charge $+1$,
and two Higgsinos $\wt{H}_{u,d}^0$
and two adjoint fermions $\chi_{a=1,2}^0$
of $R$-charge $-1$.
Among them,
we are interested in the lightest mass eigenstate $\psi$,
which is called ``pseudo-Goldstino''.
It can get a mass only from 
explicit $R$-symmetry breaking \cite{Bertolini:2011tw,Morita:2012kh},
which we can neglect for the present purpose
and so we treat it as a massless fermion.
%
%
In the absence of the adjoint Yukawa terms 
\eq{eq:W:adjoint}, 
the pseudo-Goldstino is a mixture of $\wt{R}_{I=0,u,d}^0$;
in terms of the polar angles \eq{eq:angle:def},
\begin{align} 
\psi
 &= \wt{R}_0\cos\theta
   +\left(\wt{R}^0_u\cos\varphi+\wt{R}^0_d\sin\varphi
    \right)\sin\theta
\ \equiv
    \sum_{I=0,u,d} U_{\psi I}\wt{R}^0_I 
    \ .
 \label{eq:goldstino:state}
\end{align}
The angle $\theta$ is the mixing angle between 
$SU(2)$-singlet $\wt{R}_0$ and doublets $\wt{R}_{u,d}^0$.
Note that 
the parameter $\mu$ defined in \Eq{eq:angle:def}
can be regarded as a representative scale of Higgsino masses.

In passing,
it is interesting to note that
the pseudo-Goldstino $\psi$ can have nonzero gaugino components 
in some cases.
First, the inclusion of explicit $R$-symmetry breaking
can give a tiny gaugino component \cite{Morita:2012kh}.
In addition,
the adjoint Yukawa couplings $\lambda_{ai}$
in \Eq{eq:W:adjoint} give 
mixing terms between the $R$-charged Higgsinos $\wt{R}_{i=u,d}$ 
and the adjoint fermions $\chi_{a=1,2}$,
\begin{align} 
{}-\mathcal{L}_{\psi \chi}
\ = \sum_{a=1,2}\sum_{i=u,d}  \lambda_{ai}v_i\wt{R}_i^0\chi_a^0
\ =\ \lambda^u_S v_u \wt{R}^0_u \wt{S}   +  \lambda^d_S v_d \wt{R}^0_d \wt{S}
    +\lambda^u_T v_u \wt{R}^0_u \wt{T}^0 + \lambda^d_T v_d \wt{R}^0_d \wt{T}^0
     \ .
\end{align}
where $\chi_1^0=\wt{S}$ and $\chi_2^0=\wt{T}^0$.
Diagonalizing it gives 
$\psi$ a gaugino component of $\order{\lambda_{ai}v_i/m_D}$. 
Although this can be neglected in our later analysis,
it may be relevant for other purposes,
\eg, cosmological implications of the pseudo-Goldstino.


Next we turn to the pseudo-moduli $\phi$,
the lightest mass eigenstate in the $R$-charged scalars $R_{0,u,d}^0$.
Essentially it is a fluctuation 
along the 'flat' direction \eq{eq:moduli:state},
but in the presence of soft terms 
as well as $D$-terms and adjoint Yukawa terms,
it can deviate from that direction.
That is, if we write the pseudo-moduli state as
\begin{align}
\phi
\ \equiv
    \sum_{I=0,u,d} U_{\phi I} R^0_I
    \ ,
 \label{eq:pmoduli:state}
\end{align}
the mixing angles $U_{\phi I}$ are in general
different from $U_{\psi I}$ in \Eq{eq:goldstino:state}.
Accordingly,
important for the pseudo-moduli mass $m^2_{\phi}$
are the soft scalar masses of the $R$-charged scalars $R_{0,u,d}$.
Later we will consider the case in which 
the doublets $R_{u,d}$ have a common soft mass $m^2_{R}$,
while the singlet $R_0$ can have a different mass $m^2_{R_0}$,
\begin{align}
V_{\mathrm{soft},R}
\ =\ m^2_{R_0}\abs{R_0}^2
    +m^2_{R}\sum_{i=u,d}\abs{R_i}^2
     \ .
\label{eq:V:soft:R}
\end{align}
When $m_{R_0}^2=m_R^2$,
the pseudo-moduli can be made heavy, $m^2_\phi=m^2_\psi+m^2_R$,
with the same mixing angles $U_{\phi I}=U_{\psi I}$,
as in the pseudo-Goldstino state \eq{eq:goldstino:state}.

It may be plausible that
the singlet $R_0$ does not have a soft mass
as is often the case in gauge mediation of SUSY breaking.
In this case, $m_{R_0}^2=0$,
the pseudo-moduli gets a mass $m^2_\phi$ 
through the mixing to the doublets $R_{u,d}$,
so that $m^2_\phi$ shows some interesting behavior.
For later reference, let us elucidate the $m^2_{R_0}=0$ case
in the isospin symmetric case, \ie,
by assuming $\tan\beta=1$ and $U_{\phi u}=U_{\phi d}$
so that $\phi$ lies along
$D$-flat direction of $R_{u,d}$.
When the soft mass $m_R$ can be regarded as a small perturbation,
the mass eigenvalue behaves like
\begin{align}
m^2_\phi
\ \sim\ m^2_R\sin^2\theta
\ \sim\ m^2_R\,\frac{\lambda^2 v^2}{\lambda^2 v^2+\mu^2}
        \ .
\label{eq:moduli:mass:mu:small}
\end{align}
Conversely,
when the soft mass parameter $m_R$ as well as $\mu$ 
are much greater than $\lambda v$,
the singlet $R_0$ dominates the pseudo-moduli,
so that its mass eigenvalue behaves like
\begin{align}
m^2_\phi
\ \sim\ \lambda^2 v^2
        \frac{m^2_R}{m^2_R+\mu^2}
        \ .
\label{eq:moduli:mass:mu:large}
\end{align}
In these cases,
the pseudo-moduli $\phi$ can be well lighter than
the SM-like Higgs boson.
We will further examine the properties of the pseudo-moduli
in \S{\ref{sec:decay}}.


\section{Perturbative Unification and Lightest Higgs Mass}
\label{sec:triviality}

In this section, we examine gauge coupling unification 
and triviality 
bound on the lightest Higgs mass 
in the present $R$-symmetric model.
Since the model is a kind of 
singlet extension of the minimal $R$-symmetric model,
the most important for reproducing the Higgs mass of $125\nunits{GeV}$
is the singlet Yukawa interaction $\lambda R_0 H_u H_d$,
which is subject 
to the triviality (perturbativity) constraint\,\footnote{
An alternative possibility is 
to implement the fat Higgs idea \cite{Harnik:2003rs}
in an $R$-symmetric setup,
but we do not pursue it here.
}
as in the conventional singlet extensions of the MSSM.
To discuss the triviality bound,
we need a ``UV completion'' of Dirac gaugino model.
Here we adopt a minimal model 
that is consistent with perturbative unification,
although we do not discuss its embedding into
any concrete realizations of grand unified theories.

\subsection{Gauge Coupling Unification} 
\label{subsec:unification}

Gauge coupling unification is not automatic
in models with Dirac gaugino
since we introduce 
adjoint chiral multiplets $\mathcal{A}_{a=3,2}$ 
which contribute to gauge beta functions $\beta_{g_a}$.
In addition, $R$-charged Higgs doublets $R_{u,d}$
are introduced.
As was noted 
by many authors \cite{Fox:2002bu,Kribs:2007ac,Benakli:2014cia}, 
the simplest way to recover the unification 
is to add two pairs of $SU(3)\times SU(2)$--singlets
with $U(1)_Y$--charge $\pm 1$, $\left(E^c_{4,5},\wb{E}^c_{4,5}\right)$,
with a mass term $\ME E^c_i \wb{E}^c_i$ ($i=4,5$).
For definiteness,
we refer to these `bachelor'' states\,\footnote{
Phenomenological implications of these extra ``leptons''
would be interesting but are beyond the scope of the present paper.
}
 as ``extra leptons''.
[See Table~\ref{tab:contents}.]
The resultant extra matter content is 
consistent with $SU(3)_c\times SU(3)_L\times SU(3)_R$ 
trinification \cite{trinification1,trinification2}:
the extra matter fields,
$\mathcal{A}_a$, $R_{u,d}$ as well as
the `bachelor'' $\left(E^c_{4,5},\wb{E}^c_{4,5}\right)$,
can be embedded into an adjoint representation of $SU(3)^3$,
modulo the SM singlets. 
Accordingly
one-loop coefficients of gauge beta functions,
$\left(16\pi^2\right)dg_a/dt=b_a g_a^3$ ($a=3,2,1$), 
are shifted by the same amount, $b_a=b_a^{\mathrm{MSSM}}+3$,
where $g_1=\sqrt{5/3}\,g_Y$.

Note that
the extra contribution to gauge beta functions
makes the UV gauge couplings stronger than those in the MSSM.
This fact is important
since it relaxes triviality bound for the Higgs mass,
as we shall see shortly.
In this respect, 
the present setup is the minimal one:
adding more fields would make the gauge couplings stronger
at UV and further relax the triviality bound.
[See Ref.~\cite{Kakizaki:2013eba} 
for another nontrivial choice of extra matter content.]

\begin{figure}[btp]
  \begin{center}
		\includegraphics[scale=0.5]{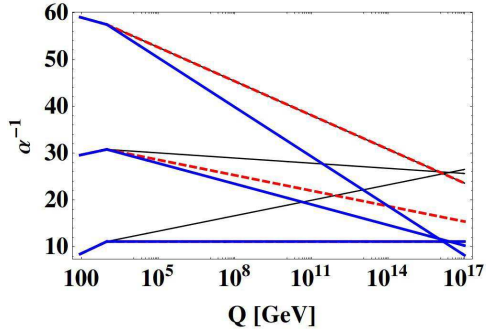}
  \end{center}
\caption{
One-loop running of gauge couplings
in the present model (thick solid)
is compared with those in the MSSM (thin solid)
and the MSSM with adjoint chiral fields added (dashed).
The vertical axis is $\alpha_a^{-1}=4\pi/g_a^2$
and horizontal axis is the renormalization scale $Q$.
}
\label{fig:GUT:1-loop}
\end{figure}

At one-loop level, gauge coupling unification is preserved,
as is depicted in Fig.~\ref{fig:GUT:1-loop},
if all the extra particles beyond the SM ones
have the common mass $\MS$ around TeV range.
Here the extra particles include
Dirac gauginos and scalar partners, 
heavier Higgses and Higgsinos, and the extra leptons, 
in addition to other SUSY particles.

Of course,
gauge coupling unification is still nontrivial 
when two-loop RGE's are used
and/or masses of the extra particles are not degenerate.
In the present work, however,
we do not intend to examine the precision of unification
to its full details, \eg, 
by taking spectrum of extra particles into account,
as was done in Ref.~\cite{Benakli:2014cia}
for the constrained minimal Dirac Gaugino SUSY SM (CMDGSSM);
Instead, we give a few examples in which unification is achieved
to the extent that is enough for estimating the bound for the Higgs mass.

In the following RG analysis,
we use two-loop RGE's summarized in \S{\ref{sec:RGanalysis}}.
In addition, we will assume that
all the extra particles are approximately degenerate around 
the scale $\MS=1\nunits{TeV}$ or $2\nunits{TeV}$,
except that
\begin{enumerate}
\item[(i)]
the extra lepton mass $\ME$ can be different from $\MS$,
\item[(ii)]
the $SU(3)$ or $SU(2)$ Dirac mass threshold,
$M_{D_3}$ or $M_{D_2}$, defined by \Eq{eq:MD:def},
can be different from $\MS$.
\end{enumerate}
Unlike the heavy Majorana gluino case,
a heavy Dirac gluino does not necessarily imply heavy squarks,
thanks to ``supersoftness'' of soft scalar masses.
This partially justifies our simplifying assumption as above.



Note that
two-loop RG evolution of gauge couplings
depends on the singlet, 
the adjoint 
and the top Yukawa couplings,
$\lambda$, $\lambda_{ai}$ and $y_t$.
Therefore the precision of unification
can be discussed only after
the correct value of the lightest Higgs mass is reproduced.
It also depends on the suppression factor $\suppressionD{D}$
for $D$-term quartic
since $\suppressionD{D}=0$, for instance, requires a larger $\lambda$,
which implies a larger two-loop effect on gauge runnings.
This being so,
further details about gauge coupling unification
at two-loop level are presented in \S{\ref{sec:unification}}.


\subsection{Triviality Bound on Singlet Yukawa Coupling}
\label{subsec:triviality}

Having discussed a UV completion of the model,
we now examine the triviality bound,
by requiring that
no coupling constant exceeds a perturbativity bound $\sqrt{4\pi}$ 
up to UV cutoff $\Lambda$,
for which we consider two cases:
lower cutoff $2.0\times 10^{16}\nunits{GeV}$ 
and higher one $1.0\times 10^{17}\nunits{GeV}$.

First let us focus on the case 
without the adjoint Yukawa couplings $\lambda_{ai}$.
Then the quantitative behaviour can be seen from one-loop RGE's 
for $\lambda$ and
the top Yukawa 
$y_t$, 
\begin{align} 
 \frac{dy_t}{dt}
\ =&\ \frac{y_t}{16\pi^2}
    \left(
       6y^2_t + \lambda^2
       -\frac{16}{3}g^2_3 - 3g^2_2 - \frac{13}{9}g^2_Y
    \right)
    \ ,
    \label{eq:yt RGE}\\
 \frac{d\lambda}{dt}
\ =&\ \frac{\lambda}{16\pi^2}\,
    \Bigl(
        4\lambda^2 + 3y^2_t - 3g^2_2 - g^2_Y
    \Bigr)
    \ ,
\label{eq:lambda:RGE}
\end{align}
where $g_Y=\sqrt{3/5}g_1$.
As we mentioned,
the UV gauge coupling constants in the present model
are larger than those in the MSSM or its singlet extensions.
It follows that
the triviality 
bound for 
$\lambda$ as well as $y_t$ is relaxed.

\begin{figure}[tbp]
  \begin{tabular}{lr}
 \begin{minipage}{0.45\hsize}
    \begin{center}
    \includegraphics[scale=0.6]{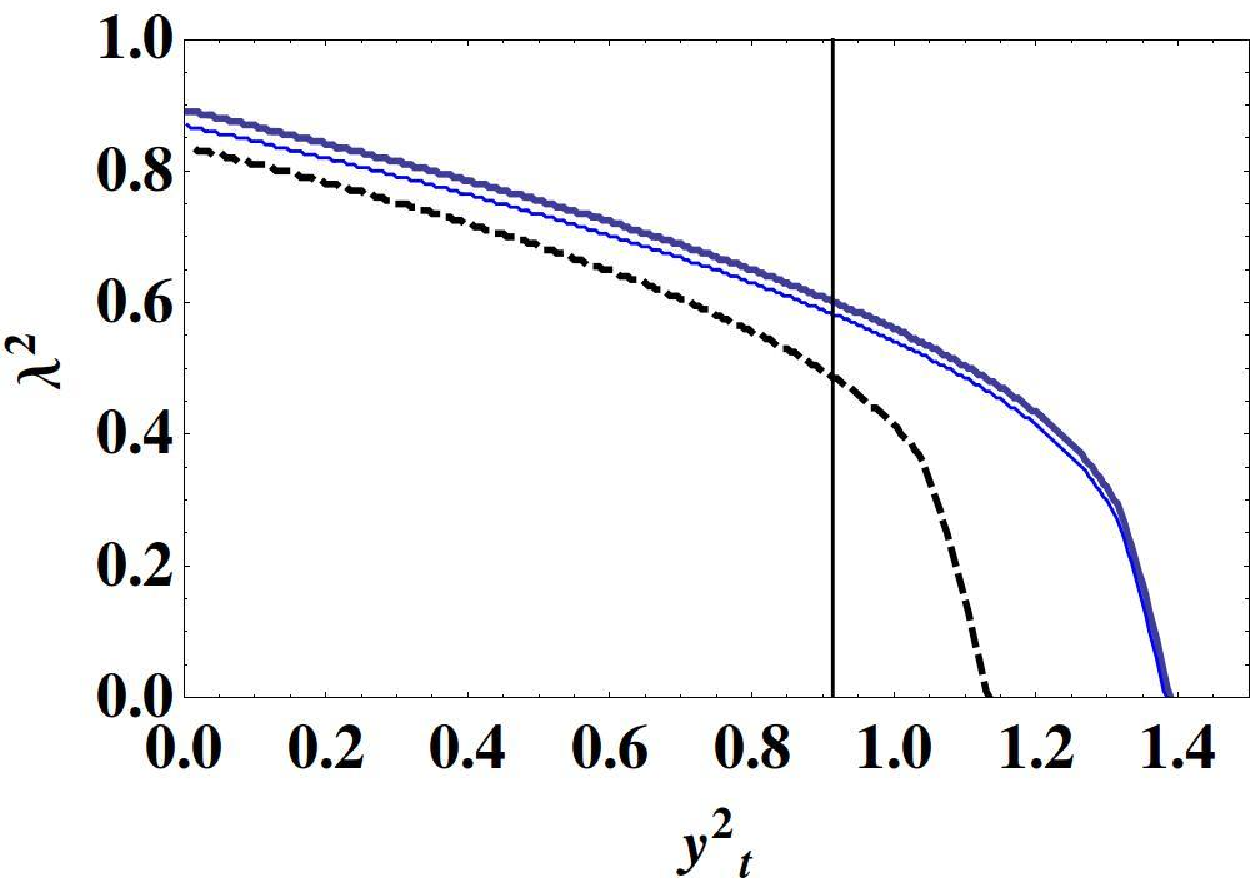}
     \end{center}
      \end{minipage}
      \hspace{3mm}
      \begin{minipage}{0.45\hsize}
       \begin{center}
      \includegraphics[scale=0.6]{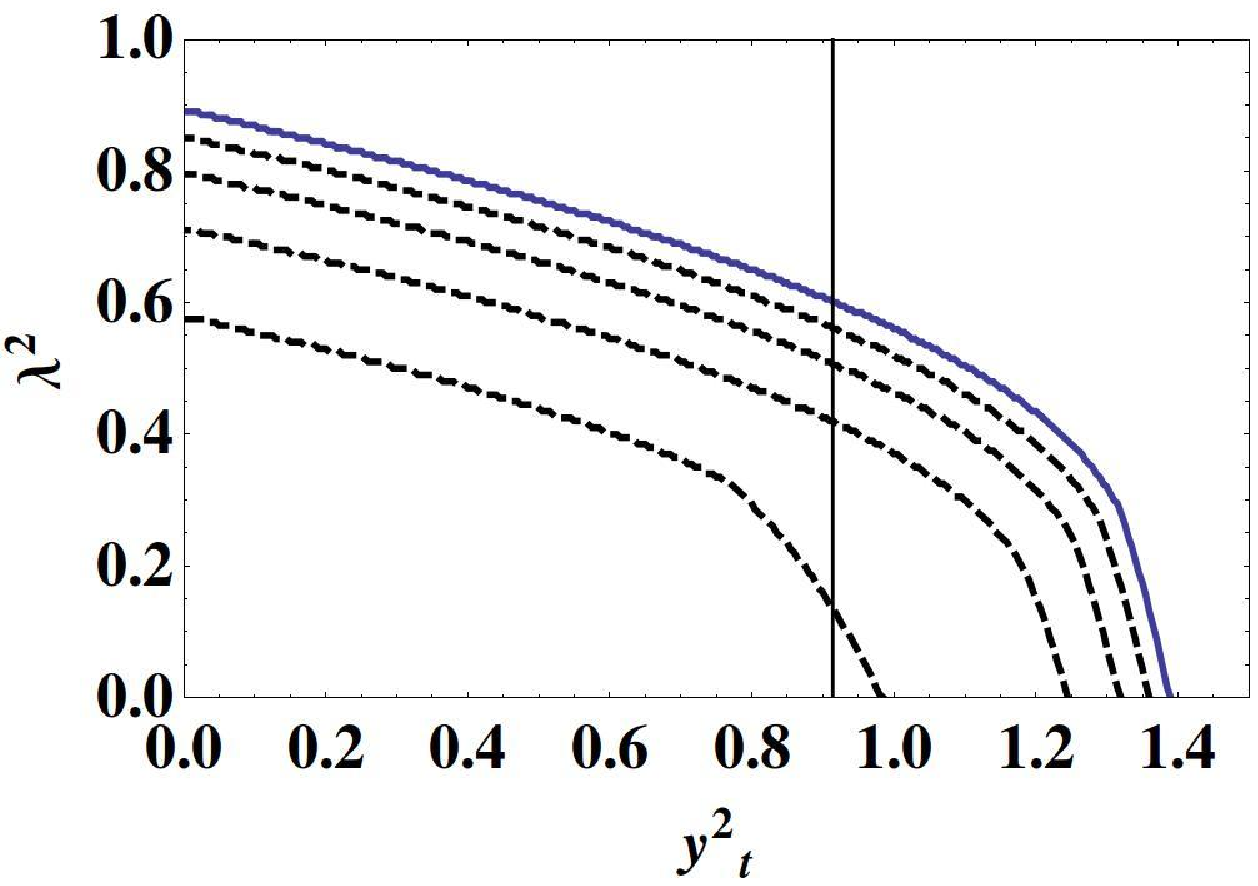}
  \end{center}
	\end{minipage}
 \end{tabular}
\caption{
Triviality bound on squared couplings $y^2_t$ and $\lambda^2$.
In the left, 
the upper solid (lower dashed) lines correspond to
the present model (nMSSM), respectively. 
The bounds are obtained
from perturbativity up to $2.0\times 10^{16}\nunits{GeV}$,
except for the thin line 
for $1.0\times 10^{17}\nunits{GeV}$.
In the right, 
the blue solid and black dashed lines correspond to
$\lambdaad=0$ and 
$\{0.2,\, 0.3,\, 0.4,\, 0.5\}$ 
from the top.
In each figure, 
all the couplings are evaluated at $\MS=1\nunits{TeV}$,
and the vertical line corresponds to $\tan\beta=2$.
 }
\label{fig:cont:pert}
\end{figure}

Figure~\ref{fig:cont:pert} shows the upper bound 
that we obtain by using two-loop RGE's. 
In the left panel,
the present case is compared with
the nMSSM-like case whose matter content is 
the same as in the singlet extension of the MSSM,
but without the cubic coupling $\kappa$ of the singlet.
Numerically,
for $\MS=1\nunits{TeV}$ and $\tan\beta=2$ ($y_t=0.95$),
the upper bound on $\lambda\fun{\MS}$
from $\Lambda=2.0\times 10^{16}\nunits{GeV}$ is $0.775$,
which is improved from $0.696$ in the nMSSM-like case,
and becomes sightly reduced to $0.762$ 
for $\Lambda=1.0\times 10^{17}\nunits{GeV}$.

We also show in the right panel of Fig.~\ref{fig:cont:pert}
how the upper bound on $\lambda$ becomes tight
when the adjoint Yukawa couplings $\lambda_{ai}$ become large.
For definiteness,
we input a common value $\lambdaad$ for $\lambda_{ai}$
at the scale $\MS=1\nunits{TeV}$ or $2\nunits{TeV}$.
Numerical values are shown 
in Table~\ref{tab:lambda:critical}.


\begin{table}[tbp]
\begin{center}
\begin{tabular}{c|c||ccccc|c}
	\makebox[17mm]{ $\Lambda\units{GeV}$} &
	\makebox[14mm]{ $\MS$ } & 
	\makebox[16mm]{ $\lambdaad=0$ } & 
	\makebox[10mm]{ $0.2$ } & 
	\makebox[10mm]{ $0.3$ } & 
	\makebox[10mm]{ $0.4$ } & 
	\makebox[10mm]{ $0.5$ } &
	\makebox[15mm]{ nMSSM }
	\\ \hline\hline
	$2\times 10^{16}$ & 
	$1\nunits{TeV}$ & 
        $\quad 0.775$ & $0.748$ & $0.711$ & $0.648$ & $0.361$ & $0.696$
	\\ 
	& 
	$2\nunits{TeV}$ & 
        $\quad 0.784$ & $0.758$ & $0.725$ & $0.660$ & $0.424$ & $0.718$ 
	\\ \hline
	$1\times 10^{17}$ & 
	$1\nunits{TeV}$ &         
        $\quad 0.762$ & $0.735$ & $0.700$ & $0.632$ & $0.265$ & $0.675$
	\\ 
	&
	$2\nunits{TeV}$ & 
	$\quad 0.771$ & $0.745$ & $0.707$ & $0.644$ & $0.339$ & $0.595$ 
	\\ \hline
\end{tabular}
\end{center}
\caption{
Upper bound of the singlet Yukawa coupling $\lambda\fun{\MS}$ 
at $\tan\beta =2$. 
}
\label{tab:lambda:critical}
\end{table}

\subsection{Triviality Bound on Higgs Mass}
\label{subsec:Higgsmass}

Now, we examine the mass of the lightest Higgs boson
and calculate its upper bound.
Here we adopt the RG approach,
and calculate the lightest Higgs mass 
by matching the present $R$-symmetric SUSY model
to a low-energy effective theory at the single scale $\MS$.
To be more precise,
starting from the input parameters summarized 
in \S{\ref{sec:inputs}},
we evolve the effective theory couplings
from the top mass scale to the matching scale $\MS$,
at which we switch to the $R$-symmetric SUSY model.
At this step, we input a value of $\tan\beta=\VEV{H_u^0}/\VEV{H_d^0}$ 
to match the top Yukawa coupling.
Then we evolve the SUSY couplings to UV region
and require the perturbativity as before.

As for the matching scale,
we take $\MS=1\nunits{TeV}$ or $2\nunits{TeV}$.
Notice that
this corresponds to a relatively low SUSY scale,
which is still consistent with the LHC bounds
thanks to ``supersafeness'' of Dirac gaugino scenario
\cite{Nojiri:2007jm,Heikinheimo:2011fk,Kribs:2012gx}.
At the same time, however,
it is quite nontrivial to reproduce the $125\nunits{GeV}$ Higgs
since 
radiative corrections from the top-stop sector are not so large;
$A$-terms are forbidden by the $R$-symmetry.



As for the low-energy effective theory,
we mainly consider the minimal SM model
by taking the decoupling limit of heavier Higgs mass eigenstates.
[We will also examine the matching
to the SM coupled with a light pseudo-moduli
in \S{\ref{subsec:match2pmoduli}}.]
Then the matching condition to the quartic Higgs potential, 
$V_{\textrm{eff}}\fun{H}=\left(\lambda_H/2\right)\abs{H}^4$,
is given by
\begin{align} 
 \lambda_H\fun{\MS}
 &=\  \frac{1}{2}\,\lambda^2\fun{\MS} \sin^2{2\beta}
    + \frac{1}{4}\,\suppressionD{D\,}g^2_Z\fun{\MS}\cos^2{2\beta}
      \ ,
\label{eq:matching:SM}
\end{align}
where $g_Z^2=g_Y^2+g_2^2$ and 
$\suppressionD{D}$ is the (common) suppression factor of $D$-term 
defined in \Eq{eq:Dterm:pot}.

\begin{figure}[tbp]
 \begin{tabular}{lr}
  \begin{minipage}{0.45\hsize}
    \begin{center}
      \includegraphics[scale=0.8]{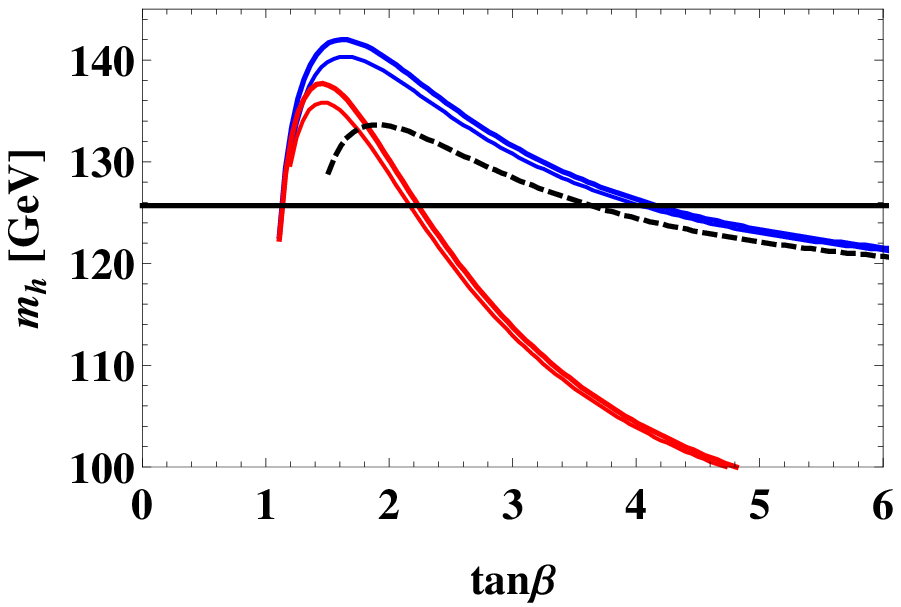}
    \end{center}
  \end{minipage}
  \hspace{3mm}
  \begin{minipage}{0.45\hsize}
    \begin{center}
	\includegraphics[scale=0.8]{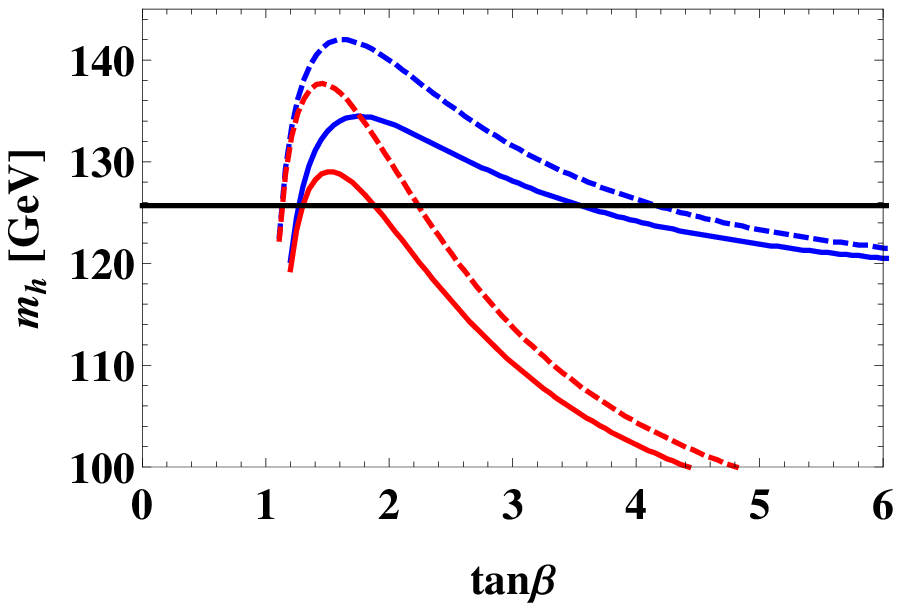}
    \end{center}
  \end{minipage}
 \end{tabular}
\begin{tabular}{lr}
  \begin{minipage}{0.45\hsize}
    \begin{center}
      \includegraphics[scale=0.8]{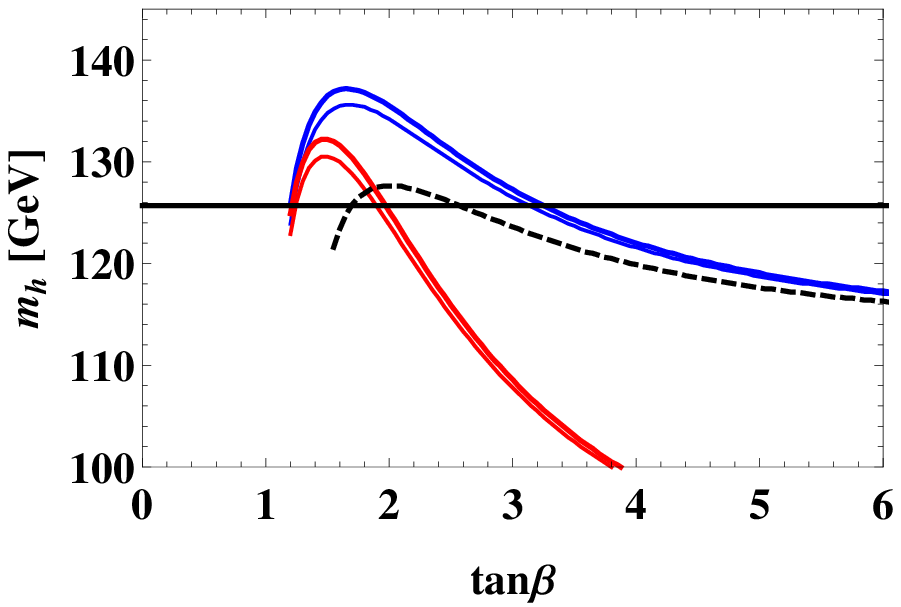}
    \end{center}
  \end{minipage}
  \hspace{3mm}
  \begin{minipage}{0.45\hsize}
    \begin{center}
	\includegraphics[scale=0.8]{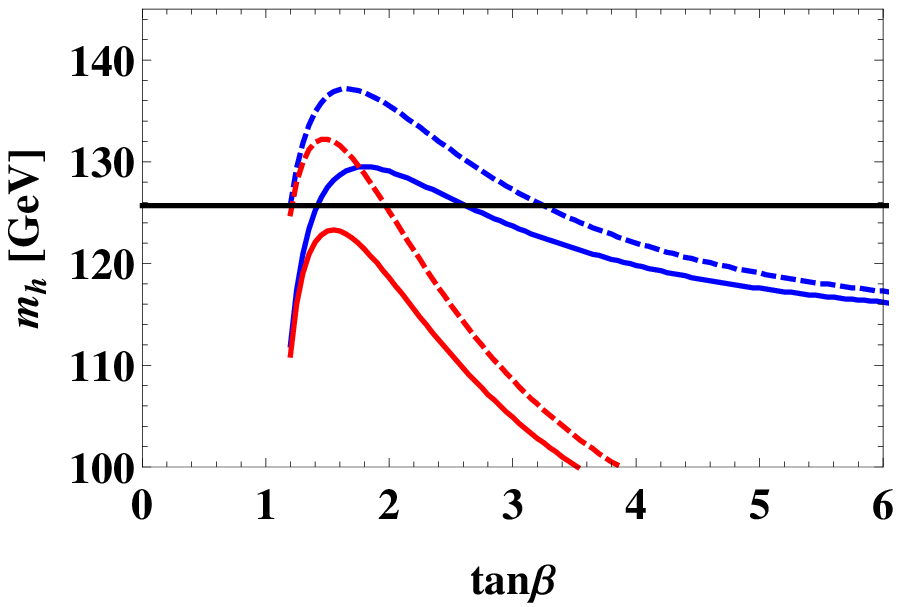}
    \end{center}
  \end{minipage}
 \end{tabular}
\caption{
Upper bounds of the lightest Higgs mass as a function of $\tan\beta$. 
The upper (lower) figures correspond to the SUSY scale
$\MS=2\nunits{TeV}$ ($\MS=1\nunits{TeV}$), respectively.
In each figure, upper blue (lower red) lines correspond to 
the unsuppressed $D$-term $\suppress{D}=1$ 
(completely suppressed $D$-term $\suppress{D}=0$), respectively.
In the left panel,
we tale $\lambdaad=0$, and
the dashed line corresponds to the nMSSM-like case.
The bounds are obtained
from perturbativity up to $\Lambda=2.0\times 10^{16}\nunits{GeV}$,
except for the thin lines 
corresponding to $\Lambda=1.0\times 10^{17}\nunits{GeV}$.
In the right panel,
we compare the $\lambdaad=0.3$ case (solid lines)
with the $\lambdaad=0$ case (dashed lines).
         }
\label{fig:Higgs:mass:bound}
\end{figure}

Figure~\ref{fig:Higgs:mass:bound} summarizes our results,
showing the upper bound on the lightest Higgs mass 
as a function of $\tan\beta$,
for various cases.
%
The left panel corresponds to the case
without adjoint Yukawa couplings.
We see that 
$125\nunits{GeV}$ Higgs mass can well be reproduced
at a small value of $\tan\beta$ around $2$--$4$.
It should be emphasized that this is possible
even without the $SU(2)\times U(1)$ quartic $D$-terms 
and with the SUSY scale as low as $1\nunits{TeV}$.

The right panel shows to what extent
the adjoint Yukawa couplings $\lambda_{ai}$ reduce the upper bounds
in the case of a common value $\lambdaad=0.3$.
We see that
$125\nunits{GeV}$ Higgs mass can be reproduced,
but requires a larger SUSY scale ($\MS=2\nunits{TeV}$)
or non-zero $D$-term ($\suppress{D}\neq0$).

\subsection{Remarks}
\label{subsec:remarks}

Some remarks are in order here.

The above result is to be compared with the nMSSM-like case, 
which typically requires a higher SUSY scale or sizable stop mixing
\cite{Ellwanger:2012ke,Ishikawa:2014owa}.
We also note that,
unlike the existing singlet extension of the MSSM,
the NMSSM or nMSSM or PQ-NMSSM \cite{Jeong:2012ma}
in which the singlet-doublet mixing can raise
the mass of the SM-like Higgs boson,
the approximate $R$-symmetry forbids such mixing
between our singlet state $R_0$ and the SM-like Higgs boson.

Another remark is that
the calculated Higgs mass may be a bit underestimated
especially for a low SUSY scale.
Actually improved two-loop calculations show
the SM-like Higgs mass receives a significant correction
in the NMSSM \cite{Goodsell:2014pla}
if the singlet is light,
and also in the minimal $R$-symmetric model (MRSSM) \cite{Diessner:2015yna}.
Nevertheless,
it is also clear that
the present model gives a sufficiently large improvement.



\begin{figure}[tbp]
 \begin{tabular}{lr}
  \begin{minipage}{0.45\hsize}
   \begin{center}
     \includegraphics[scale=0.8]{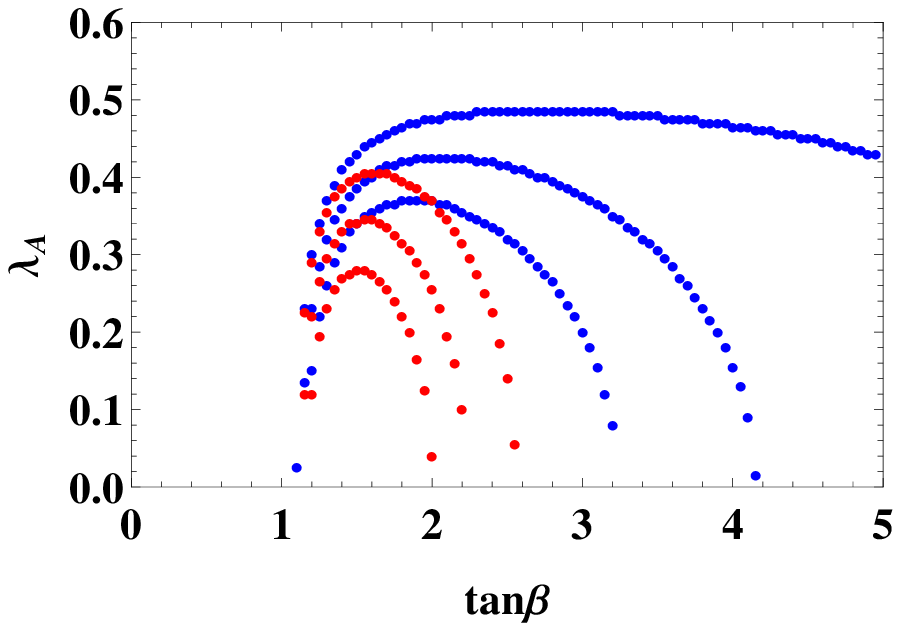} 
   \end{center}
  \end{minipage}
  \hspace{4mm}
  \begin{minipage}{0.45\hsize}
   \begin{center}
     \includegraphics[scale=0.8]{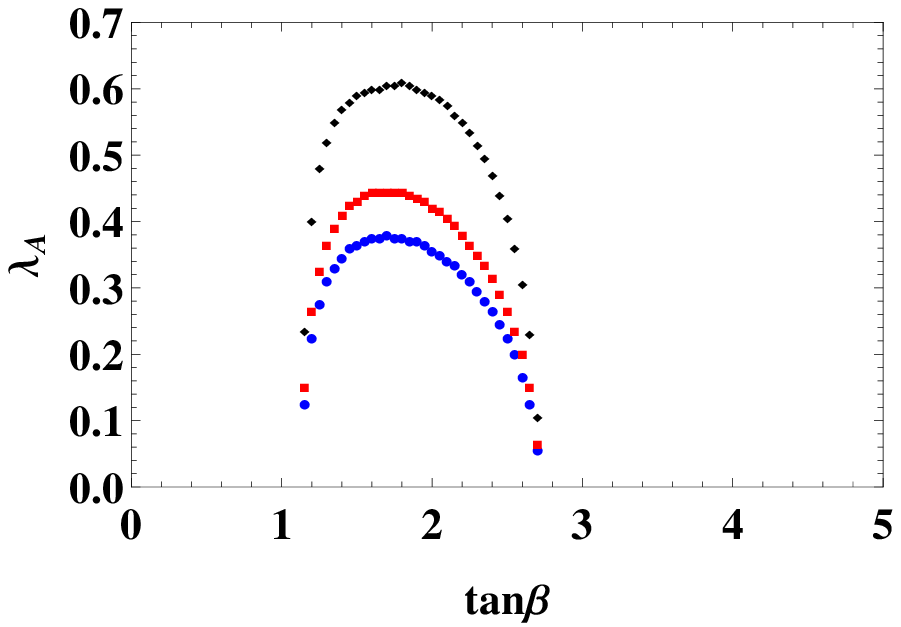}
   \end{center}
  \end{minipage}
 \end{tabular}
\caption{
The upper bounds of adjoint Yukawa couplings 
for $\MS=2\nunits{TeV}$ case.
In the left, 
the upper bounds 
corresponding to the Higgs mass 
$120$, $125$ and $130\nunits{GeV}$
are shown from the top, 
by blue dots ($\suppress{D}=1$) 
and red dots ($\suppress{D}=0$), respectively.
In the right ($\suppress{D}=0.5$),
the contours of the Higgs mass $125\nunits{GeV}$
are shown for three cases 
$(\lambda^{u,d}_S,\lambda^{u,d}_T)
=\left(\lambdaad,0\right)$,
$\left(0,\lambdaad\right)$ and
$\left(\lambdaad,\lambdaad\right)$,
from the top by black, red and blue dots, respectively.
 }
\label{fig:Higgs:mass:contour:ad}
\end{figure}

Finally let us briefly discuss to what extent
the adjoint Yukawa couplings are allowed
for realizing $125\nunits{GeV}$ Higgs mass.
Figure~\ref{fig:Higgs:mass:contour:ad} shows 
contours of the Higgs masses
in the space of $\left(\tan\beta,\lambdaad\right)$,
for a larger $\MS=2\nunits{TeV}$.
As is seen from the left panel,
if we input a common value of adjoint Yukawa couplings 
$\lambda^{u,d}_{S,T} = \lambdaad$ at the scale $\MS$,
its upper limit is around $ \lambdaad \sim 0.3\hbox{--}0.4$
depending on the $D$-term suppression factor $\suppress{D}$.
The right panel also shows that
compared with the left,
$\lambda_{T}$ can be about $15\,\%$ larger if $\lambda^{u,d}_{S}=0$,
while 
$\lambda_{S}$ can be about $38\,\%$ larger if $\lambda^{u,d}_{T}=0$.

Notice that
$\lambda=0$ corresponds to
the MRSSM, where 
the Higgs mass can be reproduced in a large $\tan\beta$ region.
In this sense,
there are two possibilities in $R$-symmetric models:
\begin{enumerate}
\item[(i)]
large $\tan\beta$ solution:
the $125\nunits{GeV}$ Higgs can 
be reproduced by the large adjoint Yukawa couplings
and corresponding radiative corrections.
\item[(ii)]
small $\tan\beta$ solution:
the $125\nunits{GeV}$ Higgs can easily be reproduced
by the large singlet Yukawa coupling
(and small adjoint couplings).
\end{enumerate}
The former requires nontrivial mass splitting
within $R$-charged Higgses and $SU(2)\times U(1)$ adjoints
so that the large adjoint coupling(s)
can play a similar role as the top Yukawa $y_t$ in the MSSM.
In our treatment,
such effects are not incorporated
since we are treating the SUSY spectrum
as a single mass threshold.

\subsection{Matching to the SM coupled with Pseudo-Moduli}
\label{subsec:match2pmoduli}


As we described in \S{\ref{subsec:pmoduli}},
the present model contains
a light degree of freedom, pseudo-moduli $\phi$, 
which is a complex scalar and can couple 
to the SM-like Higgs field.
Therefore it can affect the mass of the SM-like Higgs.
To estimate this effect,
we consider the effective theory
that contain the SM and the pseudo-moduli,
\begin{align} 
 V_{\rm{eff}}\fun{H,\phi}
 &= \frac{1}{2}\,\lambda_H \abs{H}^4 
  + \lambda_{\phi H} \abs{\phi}^2 \abs{H}^2
  + \lambda_\phi \abs{\phi}^4
    \ .
 \label{eq:CSSM}
\end{align}
It is not clear to us which field
should be identified with the moduli field $\phi$ 
in a symmetric phase of the effective theory
since 
the singlet $R_0$ can not mix with the doublets $R_{u,d}$
before the EWSB.
Recall from \S{\ref{subsec:pmoduli}}, however,
that the pseudo-moduli $\phi$ is light
especially when the Higgsino mass parameter $\mu$ is large 
and the soft mass $m_R$ is small.
In this situation,
the singlet scalar $R_0$ dominates the pseudo-moduli mass eigenstate.
So we construct the low-energy effective theory
by identifying the pseudo-moduli $\phi$ with the singlet $R_0$.

To be specific, let us assume $\mu=\order{\MS}$
and integrate out the heavy doublet fields $R_{u,d}$.
This leads to the tree-level matching condition 
\eq{eq:matching:SM} supplemented by
\begin{align} 
 \lambda_{\phi H}(\MS) 
 &=  \suppressionD{R\,}
     \lambda^2(\MS)
     \ , \qquad 
 \lambda_\phi(\MS) = 0
     \ ,
\label{eq:matching:portal}
\\
\suppressionD{R}
 &\equiv
      \frac{m^2_{R_u} }{\mu^2_u + m^2_{R_u} } \cos^2\beta
    + \frac{m^2_{R_d} }{\mu^2_d + m^2_{R_d} } \sin^2\beta
    \ ,
\label{eq:suppressionR}
\end{align}
where $m_{R_{u,d}}$ are the soft masses of $R_{u,d}$.
Here we have introduced a factor $\suppress{R}$
which represents a suppression of the ``portal'' coupling
$\lambda_{\phi H}$ at the matching scale.
This is a kind of non-decoupling effect
proportional to soft SUSY breaking parameters,
$\suppress{R}\rightarrow m^2_R/\left(\mu^2+m^2_R\right)$
for $\mu_u=\mu_d$ and $m^2_{R_u}=m^2_{R_d}$,
which is small if $m_R\ll\mu$.
On the other hand,
it can be of $\order{1}$ if $m_R\gg \mu$.
Even in this case,
\Eq{eq:moduli:mass:mu:large} shows that
the $\phi$ can remain light if $m_{R_0}=0$.
Then the ``portal'' coupling  $\lambda_{\phi H}$ is quite large at $\MS$.

For definiteness, we examine the extreme case
in which the pseudo-moduli is lighter than the SM-like Higgs boson.
Then we evolve the couplings in the effective theory \eq{eq:CSSM}
from $\MS$ down to the Higgs mass scale,
by using two-loop RGE's shown in \S{\ref{sec:RGE:CSSM}}.
Here we show one-loop parts,
which already contain the most important term:
\begin{align} 
 \beta_{\lambda_H}^{(1)}
 &=    \left. \beta_{\lambda_H}^{(1)}\right\vert_{\mathrm{SM}}
    +2 \lambda_{\phi H}^{2}   
    \ , 
\label{eq:RGE:lambdaH:oneloop}\\
 \beta_{\lambda_{\phi H}}^{(1)}
 &=  \lambda_{\phi H}
     \left(  6 \lambda_H + 4 \lambda_{\phi H}+ 2 \lambda_{\phi}   
	 + 6 y^2_t 	 -\frac{9}{2} g_{2}^{2} -\frac{15}{2}
         g_{Y}^{2} 
     \right)
     \ ,\\
 \beta_{\lambda_\phi}^{(1)}
 &=  8 \lambda_{\phi H}^{2}+ 5 \lambda_{\phi}^{2}
     \ ,
\end{align}
where the SM contributions are given by
\begin{align} 
 \left. \beta_{\lambda_H}^{(1)}\right\vert_{\mathrm{SM}}
 &=  12 \lambda^2_H +12 \lambda_H y^2_t  -12 y^4_t
    - \left(3 g_Y^2 +9 g_2^2\right) \lambda_H 
    +\frac{3}{4}g_{Y}^4 + \frac{3}{2}g_{Y}^{2} g_{2}^2 + \frac{9}{4}g_{2}^4 
    \ , \\
 \beta_{y_t}^{(1)}
 &=   y_t \left(
            \frac{9}{2} y^2_t 
            - \frac{17}{12} g_{Y}^{2} - \frac{9}{4} g_{2}^{2} -8 g_{3}^{2}
          \right) 
          \ .
					\label{eq:RGE:ytSM:oneloop}
\end{align}
\Eq{eq:RGE:lambdaH:oneloop} clearly shows that
the portal coupling $\lambda_{\phi H}$
has an effect of reducing the Higgs mass.

\begin{figure}[tbp]
 \begin{tabular}{lr}
  \begin{minipage}{0.45\hsize}
    \begin{center}
	\includegraphics[scale=0.8]{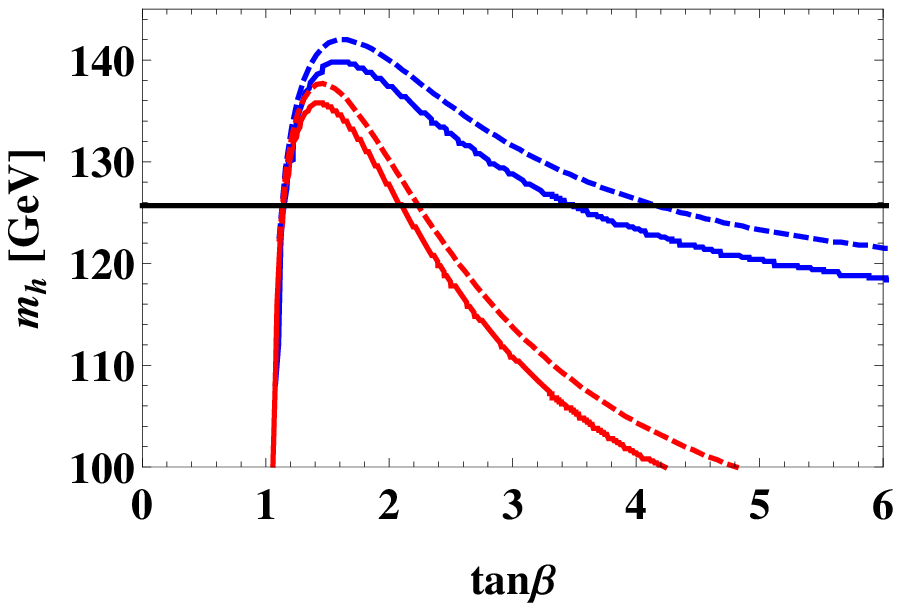}
    \end{center}
  \end{minipage}
  \hspace{4mm}
  \begin{minipage}{0.45\hsize}
    \begin{center}
	\includegraphics[scale=0.8]{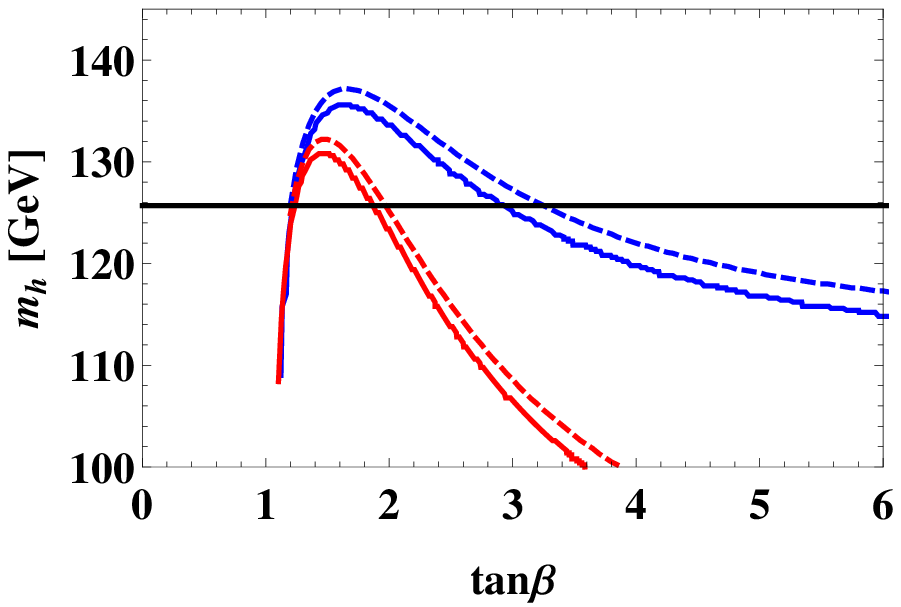}
    \end{center}
  \end{minipage}
 \end{tabular}
\caption{
Upper bounds of the lightest Higgs mass as a function of $\tan\beta$. 
In each figure,
the matching to the SM with a light moduli is assumed
in the solid lines 
with the unsuppressed matching condition $\suppress{R}=1$,
while the dashed lines correspond
to the matching to the SM ($\suppress{R}=0$),
which are the same as the solid ones in 
the left panel of Fig.~\ref{fig:Higgs:mass:bound}.
The right (left) panel corresponds to the SUSY scale
$\MS=2\nunits{TeV}$ ($\MS=1\nunits{TeV}$), 
and as before,the blue and red lines correspond to
$\suppressionD{D}=1$ case and $\suppressionD{D}=0$ case,
respectively.
 }
\label{fig:Higgs:mass:pmoduli}
\end{figure}

Figure~\ref{fig:Higgs:mass:pmoduli} shows our result.
We plot the $\suppress{R}=1$ case by the solid line,
which is to be compared with the $\suppress{R}=0$
shown by the dashed line.
Numerically,
if we compare these two cases
by the peak values of the calculated Higgs mass,
then the reduction is of $1.17$\,\% ($1.57$\,\%)
for $\MS=1\nunits{TeV}$ ($2\nunits{TeV}$)
for unsuppressed $D$-term case $\suppress{D}=1$,
while it is reduced by $1.07$\,\% ($1.40$\,\%)
for completely suppressed $D$-term $\suppress{D}=0$.
We see that 
the inclusion of the pseudo-moduli
in the low-energy effective theory
does reduce the Higgs mass,
but such effect is only at most of a few percents.
As is expected,
such reduction becomes important
when the matching scale becomes large.

\section{Constraints from Invisible Decays}
\label{sec:decay}

In this section, 
we examine how the present model is constrained from 
invisible decays of the Higgs boson~\cite{Belanger:2013kya}
as well as the $Z$ boson. 
To calculate the mass and coupling of the pseudo-moduli,
we use as an input value at $\MS=1\nunits{TeV}$
of the singlet Yukawa coupling $\lambda =0.73$ 
that corresponds, at $\tan\beta=2$,
to the lightest Higgs mass $125\nunits{GeV}$. 
No radiative correction is taken into account in this section.

\subsection{Mass  of Pseudo-Moduli}

The present singlet extension of the minimal $R$-symmetric model 
contains a light scalar 
corresponding to a fluctuation along the ``pseudo-moduli'' direction.
As was described in \S{\ref{subsec:pmoduli}},
the pseudo-moduli $\phi$ can receive a mass $m^2_\phi$
from various sources
in the scalar potential \eq{eq:V}:
the $D$-term, adjoint Yukawa terms, as well as 
soft scalar masses of $R$-charged Higgses:
\begin{align}
m_\phi^2
\ =\ m^2_{\phi,F}+m^2_{\phi,D}+m^2_{\phi,A}+m^2_{\phi,\mathrm{soft}}
     \ .
\label{eq:moduli:mass}
\end{align}
For soft term,
we assume the form \eq{eq:V:soft:R},
\begin{align}
V_{\mathrm{soft},R}
\ =\ m^2_{R_0}\abs{R_0}^2
    +m^2_{R}\sum_{i=u,d}\abs{R_i}^2
     \ ,
\end{align}
where the soft mass $m_{R_0}$ of the singlet $R_0$
can be different from 
the common soft mass $m_{R}$ of the doublets $R_{u,d}$.
When all the $R$-charged Higgses 
have a universal soft mass $m^2_{R_0}=m^2_R(=m^2)$, 
the pseudo-moduli has a mass equal to $m^2_\phi=m^2$
and its mixing angles coincide with those of 
the pseudo-Goldstino states \eq{eq:moduli:state},
$U_{\phi I}=U_{\psi I}$ ($I=0,u,d$).

\begin{figure}[tbp]
 \begin{tabular}{cc}
  \begin{minipage}{0.45\hsize}
	\includegraphics[scale=0.8]{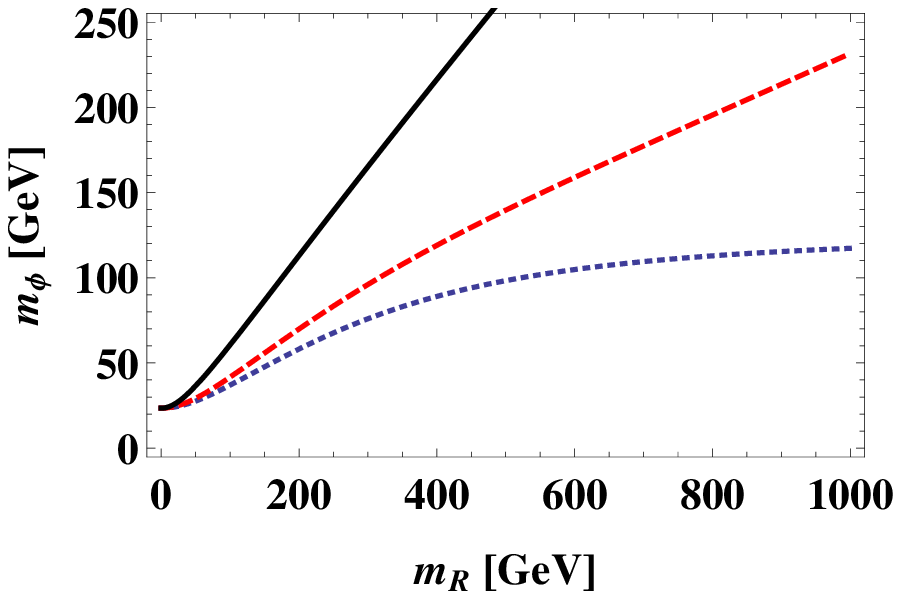} 
  \end{minipage}
  \hspace{4mm}
  \begin{minipage}{0.45\hsize}
	\includegraphics[scale=0.8]{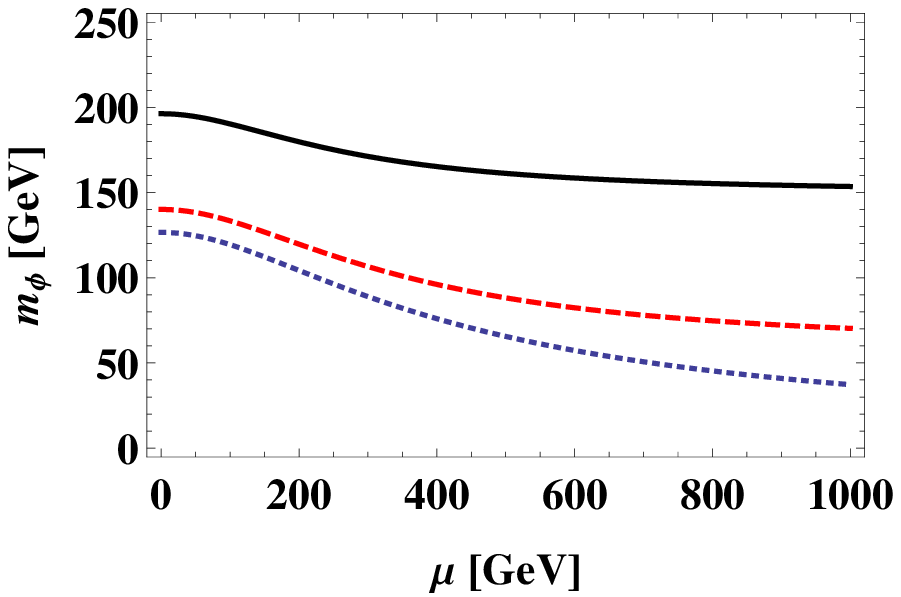}
  \end{minipage}
 \end{tabular}
\caption{
The pseudo-moduli mass $m_\phi$ as a function of
the averaged $\mu$ parameter (left) 
and the common doublet soft mass $m_R$ (right).
We take $\lambda =0.73$ and $\tan\beta=2$; 
$\mu=400\nunits{GeV}$ on the left,
while $m_R=300\nunits{GeV}$ on the right.
%
For the singlet soft mass $m_{R_0}$,
three cases $m_{R_0}/m_R=0$, $0.2$, and $0.5$ are shown
by blue dotted, red dashed and solid lines, respectively.
%
	}
\label{fig:moduli:mass}
\end{figure}

We are particularly interested in the limiting case,
$m^2_{R_0}\ll m^2_R$.
Figure~\ref{fig:moduli:mass} shows 
the mass eigenvalue of the pseudo-moduli state $\phi$,
in three particular cases,
$m^2_{R_0}=0$, $m^2_{R_0}=0.2\,m^2_R$ and $m^2_{R_0}=0.5\,m^2_R$.
From the figure, we see the following:
\begin{itemize}
\item
For a fixed value of $\mu$,
the pseudo-moduli mass $m_\phi$ increases
by increasing the soft mass $m_R$ of $R_{u,d}$,
and approaches the singlet mass $m_{R_0}$
in the limit of large $m_R$.
In particular, if $m_{R_0}=0$,
it approaches 
the limiting value $\lambda v$, 
which is $126\nunits{GeV}$ for $\lambda=0.73$.
\item
For a fixed value of $m_R$,
its mass $m_\phi$ decreases by increasing $\mu$.
This is because
the singlet component dominates 
the pseudo-moduli state
when $\mu$ is large compared to $\lambda v$. 
\end{itemize}
These behavior can be understood from our estimate 
\eqs{eq:moduli:mass:mu:small}{eq:moduli:mass:mu:large}.

\subsection{Interactions to the SM Higgs}

\newcommand{\mixing}[2]{\kappa_{{#1}{#2}}}
\renewcommand{\mixing}[2]{\delta_{{#1}}m_{#2}}



Next we calculate the trilinear coupling $g_{h\phi\phi}$
of the lightest Higgs boson $h$ to the pseudo-moduli pair,
which also takes the form
\begin{align} 
g_{h\phi\phi}
\ =\ g_{h\phi\phi}^{(F)}+g_{h\phi\phi}^{(D)}+g_{h\phi\phi}^{(A)}
     \ .
\end{align} 
Let us take a close look at the first term
$g_{h\phi\phi}^{(F)}$ corresponding to the superpotential \eq{eq:W:Higgs}.
Substituting the usual expansion
\begin{align} 
	\begin{pmatrix}
		H_u\\
		H_d
	\end{pmatrix}
 &=
        v
	\begin{pmatrix}
		\sin\beta\\
		\cos\beta
	\end{pmatrix}
	+\frac{1}{\sqrt{2}} 
	\begin{pmatrix}
		\cos\alpha & \sin\alpha\\
		-\sin\alpha & \cos\alpha
	\end{pmatrix}
	\begin{pmatrix}
          h\\
          H'
	\end{pmatrix}
\nonumber
\end{align}
into the corresponding term $V_F$ in the scalar potential \eq{eq:V},
we obtain terms quadratic in the pseudo-moduli field $\phi$ as
\begin{align} 
V_{\phi,F}
 &=  m^2_{\phi,F}\abs{\phi}^2
     + g_{h\phi\phi}^{(F)} \abs{\phi}^2 h
     + \frac{1}{2}\,g_{hh\phi\phi}^{(F)} \abs{\phi}^2 h^2
     \ .
 \label{eq:Higgs-mod-int:F}
\end{align} 
Here the contributions to the mass and the trilinear coupling
are given by
\begin{align} 
m^2_{\phi,F}
 &=  \abs{\delta_u m_\phi}^2 + \abs{\delta_d m_\phi}^2
     \ ,
\\
%
 g_{h\phi\phi}^{(F)}
 &=  \frac{\lambda }{\sqrt{2}}
     \left[
       U^*_{0\phi}
       \Bigl(
         \delta_u m_{\phi\,}\sin\alpha - \delta_d m_{\phi\,}\cos\alpha
       \Bigr) + \mbox{c.c.}
     \right]
     \ ,
\end{align}
where
$U_{I\phi}=U^*_{\phi I}$ is 
the component of the lowest mass eigenstate $\phi$ 
in the fields $R_{I=0,u,d}$,
and we have defined
\begin{align} 
 \mixing{u}{\phi}
 \equiv
     \frac{\partial F_{H_u}}{\partial \phi}
 &=\,\mu_u U_{u\phi}-\lambda v_d U_{0\phi}
     \ , \qquad
\nonumber\\
 \mixing{d}{\phi}
 \equiv
     \frac{\partial F_{H_d}}{\partial \phi}
 &=\,\mu_d U_{d\phi}-\lambda v_u U_{0\phi}
      \ .
\end{align}
The quartic coupling $g_{hh\phi\phi}$ can be found
in a similar manner;
we omit it here since it is irrelevant for later purposes.


Now, it is important to realize that
the pseudo-moduli mixing angles $U_{I\phi}$ ($I=0,u,d$)
coincide with pseudo-Goldstino angles \eq{eq:moduli:state},
$U_{I\phi}\rightarrow U_{I\psi}$,
if we neglect the terms in the scalar potential \eq{eq:V}
other than $V_F$;
The same is true 
if the soft masses are universal in the $R$-charged Higgs sector.
In these situations, we have 
\begin{align} 
\delta_u m_{\phi}
\ ={}-\mu\sin\theta\cos\beta
      \left[
        \frac{U_{0\phi}}{\cos\theta}
        -\frac{U_{u\phi}}{\sin\theta\cos\varphi}
      \right]
 &\longrightarrow\ 0
      \ ,
\nonumber\\
\delta_d m_{\phi}
 \ ={}-\mu\sin\theta\sin\beta
      \left[
        \frac{U_{0\phi}}{\cos\theta}
        -\frac{U_{d\phi}}{\sin\theta\sin\varphi}
      \right]
 & \longrightarrow\ 0
  \ ,
\end{align}
which reflects the fact that
the pseudo-moduli $\phi$ does not get a mass from the Higgs VEV's.
It implies that 
$\phi$ decouples from the Higgs boson
in this limit.
In other words,
the trilinear interaction of a lightest Higgs and two pseudo-moduli
arises from a deviation of the pseudo-moduli eigenstate
from the would-be flat direction,
caused by the other terms in the scalar potential \eq{eq:V}.

The $D$-term contribution to the trilinear coupling 
is\,\footnote{
This contribution can take both signs,
as in the case of $D$-contributions to soft scalar masses.
}
\begin{align} 
 g_{h\phi\phi}^{(D)}
\ ={}-\suppressionD{D\,}\frac{g_Z^2 v}{2\sqrt{2}}\,
      \sin\left(\alpha+\beta\right)
      \Big(
        \abs{U_{u\phi}}^2-\abs{U_{d\phi}}^2
      \Bigr)
      \ .
\label{eq:Higgs-mod-int:D}
\end{align}
This contribution is quite small, however,
in most of the parameter space: 
When $\mu$ or $m_R$ is larger than $\lambda v\sim 126\nunits{GeV}$,
the pseudo-moduli state is dominated by the singlet component,
so that the doublet components $U_{i\phi}$ ($i=u,d$) are small.
Moreover, in the decoupling limit of the heavier Higgses, we have 
$\sin\left(\alpha+\beta\right)\rightarrow{}-\cos 2\beta
\left({}\sim 0.6\right)$,
giving another suppression
for a small $\tan\beta\left({}\sim 2\right)$.


The adjoint Yukawa terms \eq{eq:V:adjoint}
also contribute, after the EW symmetry breaking,
to the mass matrix of $R$-charged Higgses
and the trilinear coupling to the Higgs boson
\begin{align} 
 g_{h\phi\phi}^{(A)}
 = \sum_{a=S,T}
     \frac{v}{\sqrt{2}} 
     \left[ 
       \left(
           \lambda ^u_a U_{u\phi}\sin\beta 
         + \lambda ^d_a U_{d\phi}\cos\beta 
       \right)
       \left(
           \lambda ^u_a U_{u\phi} \cos\alpha
         - \lambda ^d_a U_{d\phi}\sin\alpha
       \right)^*
     + \hbox{c.c.}
     \right] 
     \ .
\label{eq:Higgs-mod-int:ad}
\end{align}
This contribution is also small 
unless $\lambdaad{}v$ is comparable to $\mu$ or $m_R$.



\begin{figure}[tbp]
			\begin{tabular}{cc}
			\begin{minipage}{0.45\hsize}
	\includegraphics[scale=0.8]{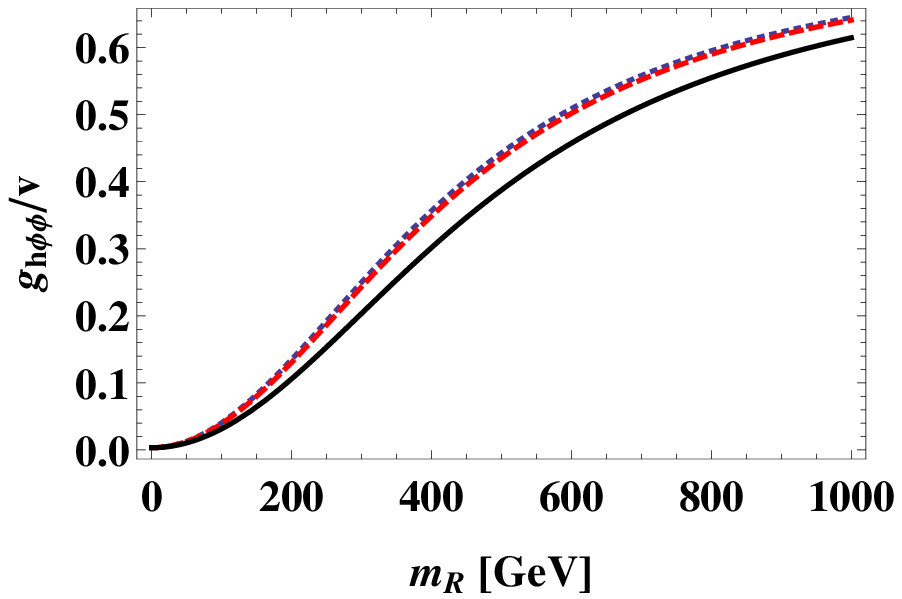}
	\end{minipage}
\hspace{4mm}
			\begin{minipage}{0.45\hsize}
	\includegraphics[scale=0.8]{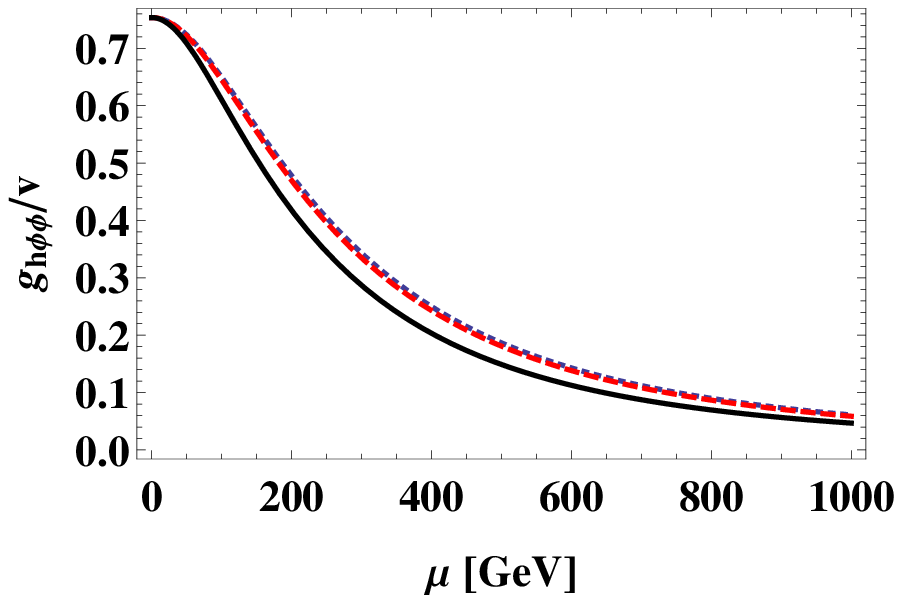}
	\end{minipage}
\end{tabular}
\caption{
The trilinear coupling $g_{h\phi\phi}$
(divided by the Higgs  VEV) 
as a function of $\mu$ and soft mass $m_R$ of $R_{u,d}$.
The parameters are the same 
as in the previous figure~\ref{fig:moduli:mass}.
	}
	\label{fig:hmodmod-int}
\end{figure}

Figure~\ref{fig:hmodmod-int} shows 
the trilinear coupling 
$g_{h\phi\phi}=g_{h\phi\phi}^{(F)}+g_{h\phi\phi}^{(D)}+g_{h\phi\phi}^{(A)}$ 
of the lightest Higgs and two pseudo-moduli.
We see that
it is a increasing function of the soft mass $m_R$
as is expected;
it also decreases as $\mu$ becomes large,
since the singlet $R_0$ dominates in the eigenstate $\phi$.

In passing, we note that
In the decoupling limit of heavier Higgses,
where $\alpha\rightarrow\beta-\pi/2$,
Eqs.~(\ref{eq:Higgs-mod-int:F}), 
\eqs{eq:Higgs-mod-int:D}{eq:Higgs-mod-int:ad}
are slightly simplified as
\begin{align} 
 g_{h\phi\phi}^{(F)}
 &\longrightarrow{}
      \sqrt{2} \lambda^2 v |U_{0\phi}|^2 
     -\frac{\lambda}{\sqrt{2}}
      \left[ U_{0\phi}^*
        \Bigl(
          \mu_u U_{u\phi}\cos\beta +\mu_d U_{d\phi} \sin\beta 
        \Bigr)
        +\hbox{c.c}
      \right]
      \ ,
\\
 g_{h\phi\phi}^{(D)}
 &\longrightarrow{}
      \sqrt{2}\,\frac{g^2_Z}{4}\,v\,\cos2\beta 
      \Bigl(
        \abs{U_{u\phi}}^2-\abs{U_{d\phi}}^2
      \Bigr)
      \ ,
\\
 g_{h\phi\phi}^{(A)}
 &\longrightarrow{}
      \sqrt{2}
      \sum_{a=S,T}
      v
      \left|
        \lambda ^u_a U_{u\phi} \sin\beta 
        + \lambda ^d_a U_{d\phi} \cos\beta 
      \right|^2
      \ .
\end{align}



\subsection{Bounds from Invisible Decays}

Having calculated the mass and the interaction
of the pseudo-moduli,
we now examine the constraints from the invisible decays.

First let us briefly discuss the constraint 
from invisible decay of the Z boson.
Since we are supposing that explicit $R$ symmetry breaking
is very small \cite{Morita:2012kh},
we treat the pseudo-Goldstino $\psi$ as massless
in the following.

The $Z$ coupling of the pseudo-Goldstinos 
comes from the neutral current
\begin{align} 
  \left.J_{Z}^\mu\right\vert_{\wt{R}}
\ =\ 
        \frac{g_Z}{2}\left[
           \wt{R}^{0\dagger} _d \wb{\sigma}^\mu \wt{R}_d^0
          -\wt{R}^{0\dagger}_u \wb{\sigma}^\mu \wt{R}_u^0
        \right]
\ ={}- \frac{g_Z}{2}\,
        \psi^\dagger \wb{\sigma}^\mu\psi\,
        \sin^2\theta \cos{2\varphi}
        +\cdots
        \ , 
\label{eq:Zpsipsi:int}
\end{align}
with the mixing angles $\theta$ and $\varphi$ 
defined by \Eq{eq:angle:def}.
The decay width of $Z\to \psi\psi^\dagger$ is then
\begin{align} 
 \Gamma_{Z\to \psi\psi}
 \ =\ \frac{g^2_Z m_Z}{96 \pi}
       \left(\sin^2\theta \cos{2\varphi}\right)^2 
       \ .
\label{eq:Zpsipsi:width}
\end{align}
Requiring that this is less than 
the error of the measured value~\cite{Agashe:2014kda} of the total width 
$ \Gamma _Z\ =\ 2.4952\pm0.0023\nunits{GeV}$,
we obtain the bound
\begin{align} 
  \abs{\sin^2\theta \cos{2\varphi}}\ \leq\ 0.0988
  \ .
\end{align}

\begin{figure}[btp]
  \begin{center}
		\includegraphics[scale=0.8]{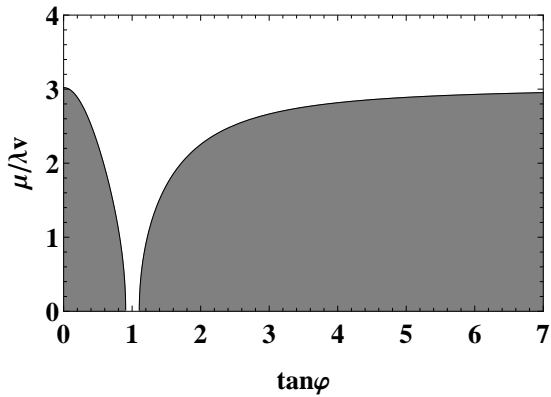}
  \end{center}
\caption{
The constraint from invisible decay of the Z boson.
The gray region is excluded. 
The vertical axis is the mixing angle $\cot\theta=\mu/\lambda{}v$ 
between the singlet/doublet components of 
$R$-charged Higgsinos, $\wt{R}_0$ and $\wt{R}_{u,d}^0$,
while the horizontal axis is the angle $\tan\varphi$ 
within $\wt{R}_{u,d}^0$.
         }
\label{fig:zinv2}
\end{figure}

The bound is depicted in Fig.~\ref{fig:zinv2}.
Recall that 
$\tan\theta=\lambda{}v/\mu$ is the mixing angle 
between the singlet and doublet components of $R$-charged Higgsinos,
$\wt{R}_0$ and $\wt{R}_{u,d}^0$.
For a generic value of $\tan\varphi$, 
which is the angle between $\wt{R}_u$ and $\wt{R}_d$,
a smaller value of $\mu$ implies that
the pseudo-Goldstino $\psi$ has a larger doublet component
and thus more strongly couples to the $Z$ boson.
We see that 
the constraint can easily be satisfied
if the averaged Higgsino mass parameter $\mu$ 
is larger than $\lambda v$:
For $\lambda=0.73$, we get a lower bound $\mu \gtrsim 285\nunits{GeV}$.
Then the chargino mass bounds from direct searches are also satisfied.
We add that
the pseudo-Goldstino becomes almost decoupled from the $Z$ boson
for $\tan\varphi=\left(\mu_u/\mu_d\right)\tan\beta \approx 1$,
or equivalently,
for $\tan\beta\approx\mu_d/\mu_u$
as is the case if the soft masses of $H_{u,d}$ are small.

The $Z$ boson can decay also into
a pair of pseudo-moduli $\phi$
if the latter is lighter than $m_Z/2$.
This corresponds to
$m_R\lsim 150\nunits{GeV}$ for $\mu=400\nunits{GeV}$ and
$\mu\gsim 700\nunits{GeV}$ for $m_R=300\nunits{GeV}$,
as is seen from Fig.~\ref{fig:moduli:mass}. 
The partial decay width for $Z\to \phi\phi^\dagger$
\begin{align} 
 \Gamma_{Z\to \phi\phi} 
\  =\ \frac{g^2_Z m_Z}{96 \pi}
       \Bigl(\abs{U_{u\phi}}^2-\abs{U_{d\phi}}^2\Bigr)^2 
       \left(1-\frac{4m^2_\phi}{m^2_Z}\right)^{3/2} 
\end{align}
approaches that of $Z\to\psi\psi^\dagger$ 
in the limit of small $m_R$,
but becomes negligible if 
the soft mass $m_R$ is as large as $100\nunits{GeV}$.



Next we discuss the invisible width of the Higgs boson.
If the pseudo-moduli $\phi$ is lighter than $m_h/2$,
the Higgs boson can decay into a pair of $\phi$.
The decay width for $h\to\phi\phi^\dagger$ is given 
in terms of the trilinear coupling $g_{h\phi\phi}$ 
and the pseudo-moduli mass $m_\phi$
by\,\footnote{
The decay $h\to\psi\wb{\psi}$ is negligible
since 
it is proportional to $m_\psi$
(modulo a loop-suppressed contribution)
and $m_\psi$ vanishes in the $R$-symmetric limit.
}
\begin{align} 
 \Gamma _{h\to\phi\phi}
\ =\ 
     \frac{g^2_{h\phi\phi}}{16\pi m_h}\,
     \sqrt{1-\frac{4m^2_\phi}{m^2_h}}
     \ .
\end{align}
%
We require that 
this width should be smaller 
than the partial width of $h \to b\wb{b}$ in the SM,
\begin{align} 
 \Gamma_{h\rightarrow\phi\phi}
\ <\ \Gamma_{h\rightarrow bb} ^{\mathrm{SM}}
\ \approx\ 2.34\times 10^{-3}\nunits{GeV}
        \ ,
\end{align}
where $\Gamma^{\mathrm{SM}} _h=4.07\times 10^{-3} \nunits{GeV}$
and $\mathrm{Br}\!\left({h \to b\wb{b}}\right)=57.7 \% $
are the SM predictions \cite{Agashe:2014kda}.

\begin{figure}[tbp]
 \begin{tabular}{cc}
  \begin{minipage}{0.45\hsize}
	\includegraphics[scale=0.5]{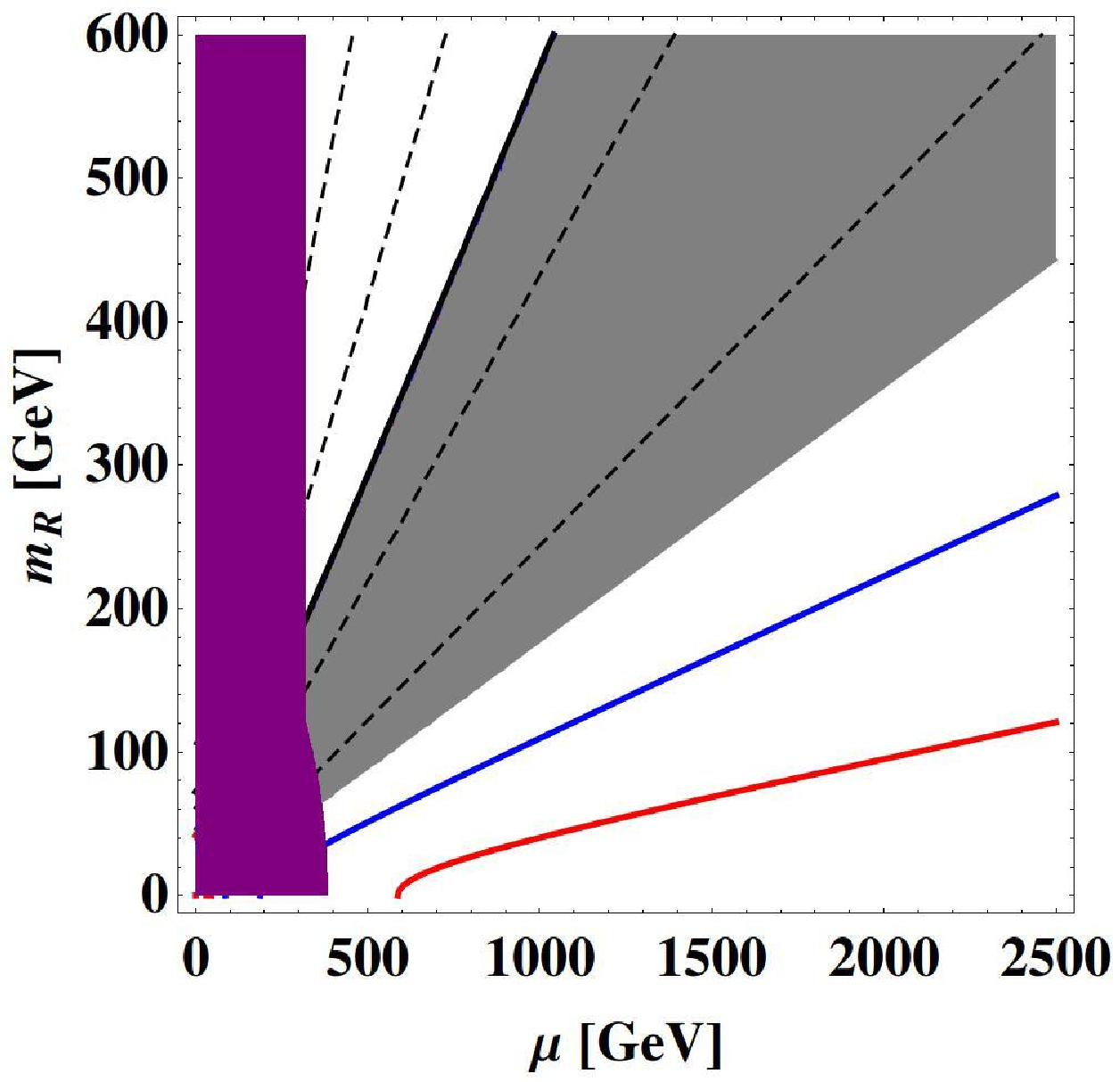}
  \end{minipage}
  \hspace{4mm}
  \begin{minipage}{0.45\hsize}
	\includegraphics[scale=0.5]{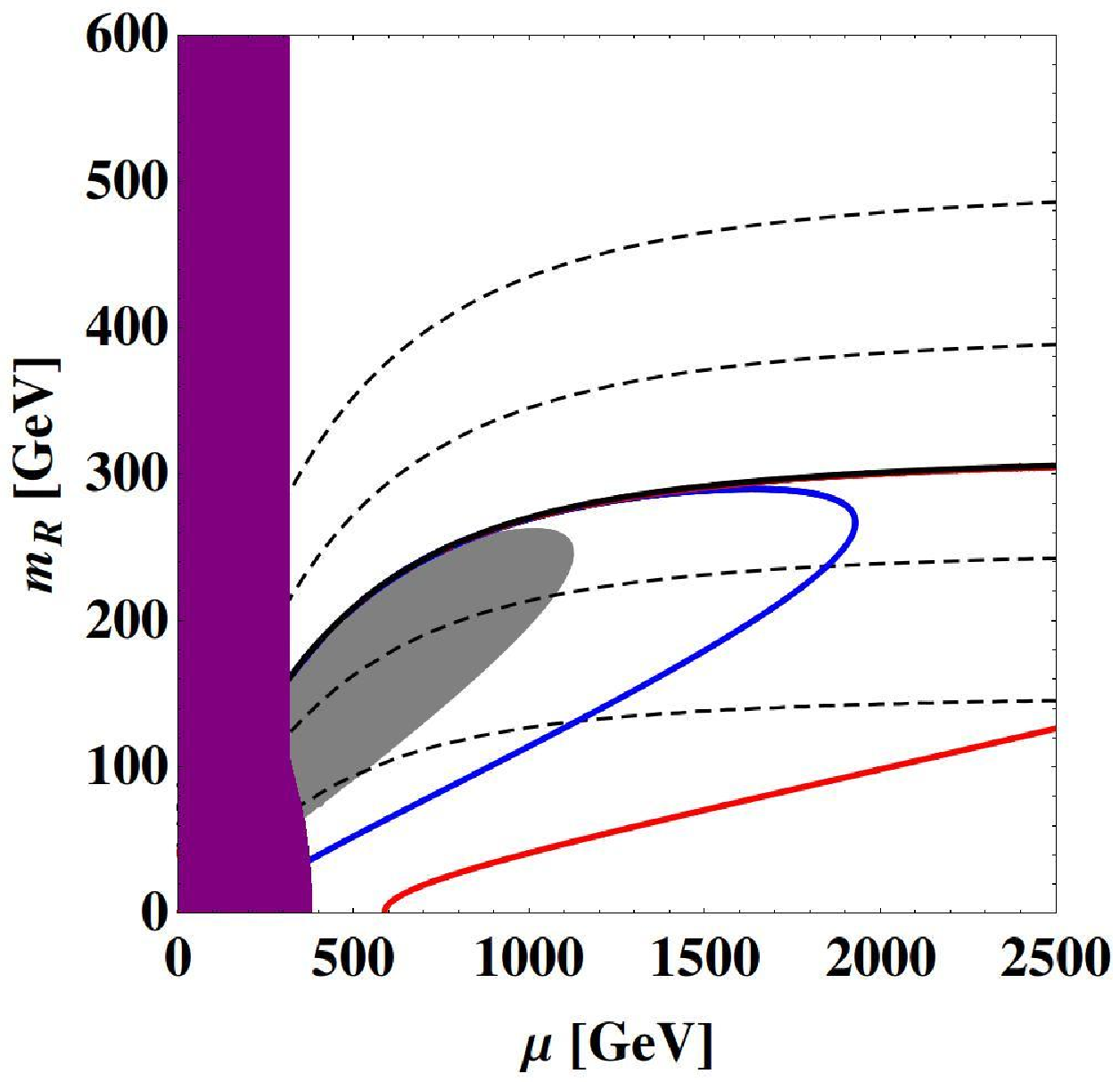}
  \end{minipage}
 \end{tabular}
\caption{
The constraints from invisible $Z$ and Higgs decays
in $\left(\mu,m_R\right)$ plane.
The input parameters are $\tan\beta=2$, $\tan\varphi=2.5$,
$\suppressionD{D}=0.5$ and $\lambda =0.73$ at $\MS=1\nunits{TeV}$.
In the left panel, we take $m_{R_0}=0$ 
while $m_{R_0}=0.2\, m_{R}$ in the right.
The purple and gray regions are excluded 
by invisible decay of the $Z$ and Higgs, respectively.
The black solid contour corresponds to $m_\phi=m_h/2$,
while the dashed contours correspond to
$m_\phi=\{30, 50, 80, 100\}\nunits{GeV}$, respectively, from the bottom.
Future reaches of 
the LHC ($\mathrm{Br}=0.09$) and ILC ($\mathrm{Br}=0.004$)
are also shown by the blue and red lines, respectively.
	}
\label{fig:Higgs:inv:decay} 
\end{figure}

Figure~\ref{fig:Higgs:inv:decay} shows
the resulting bound on parameters $\mu$ and $m_R$. 
The left panel corresponds to the $m_{R_0}=0$ case,
in which an excluded region appears
once the pseudo-moduli mass $m_\phi$ (shown by the dashed lines)
is smaller than $m_h/2$.
Remarkably and unlike a generic expectation,
the constraint becomes weak and disappears
as the pseudo-moduli becomes lighter and lighter.
We also see from the right panel that
the constraint is milder 
for the singlet soft mass $m_{R_0}\neq0$.
Actually it disappears 
if the $\mu$ parameter is as large as $1\nunits{TeV}$.

In the figures,
we also show the contours of future reaches
for the branching ratio of invisible Higgs decays,
for which we quote 
$9\%$ for the LHC with $3000\nunits{fb}^{-1}$
at $\sqrt{s}=14\nunits{TeV}$ \cite{Okawa:2013hda},
and $0.4\%$ for the ILC with $1150\nunits{fb}^{-1}$ 
at $\sqrt{s}=250\nunits{GeV}$ \cite{Asner:2013psa}.
We need other constraints, 
for instance from direct searches for charginos,
to cover the whole parameter space.

\begin{figure}[tbp]
 \begin{tabular}{cc}
 \begin{minipage}{0.45\hsize}
	\includegraphics[scale=0.5]{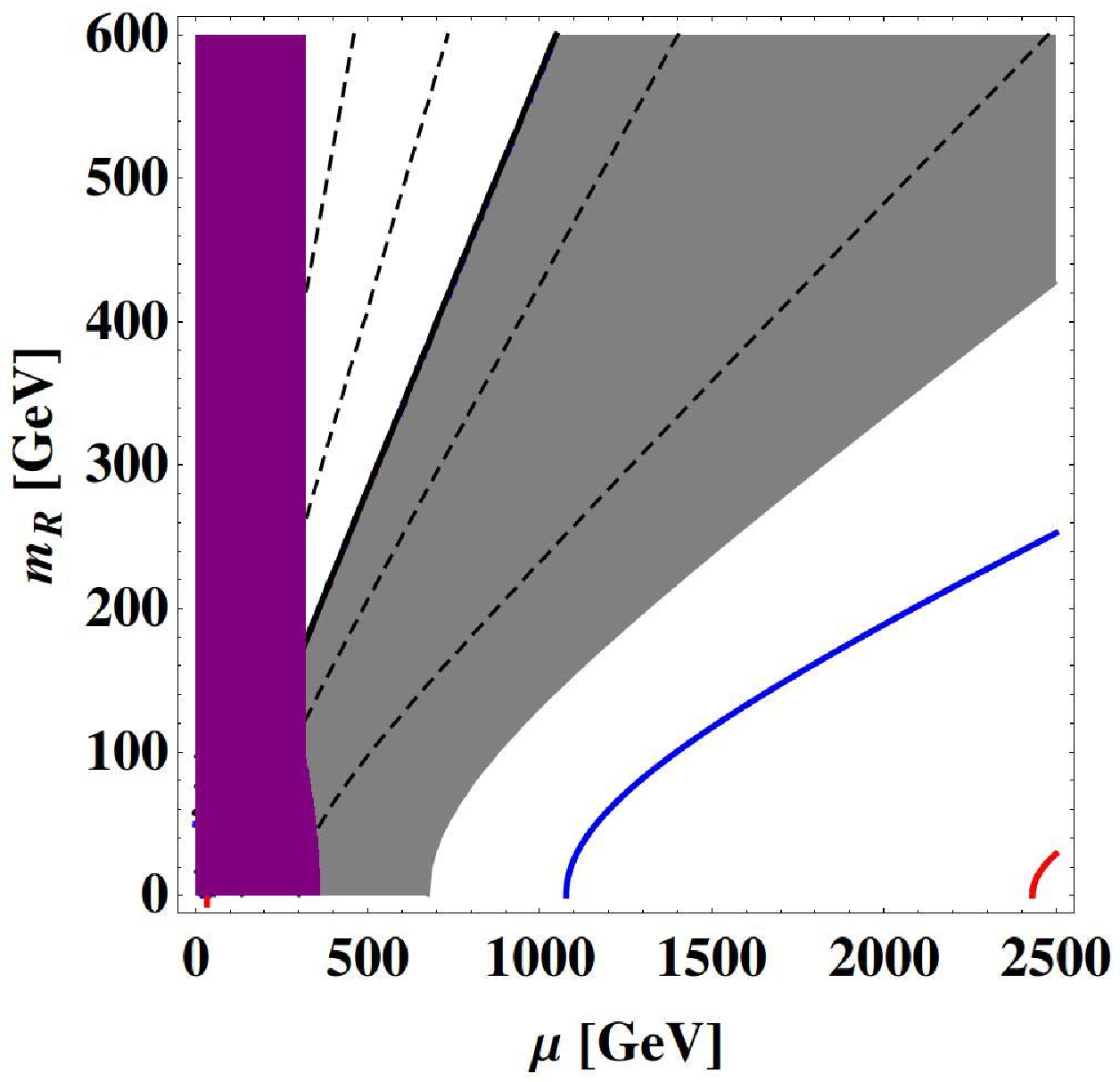}
  \end{minipage}
  \hspace{4mm}
  \begin{minipage}{0.45\hsize}
	\includegraphics[scale=0.5]{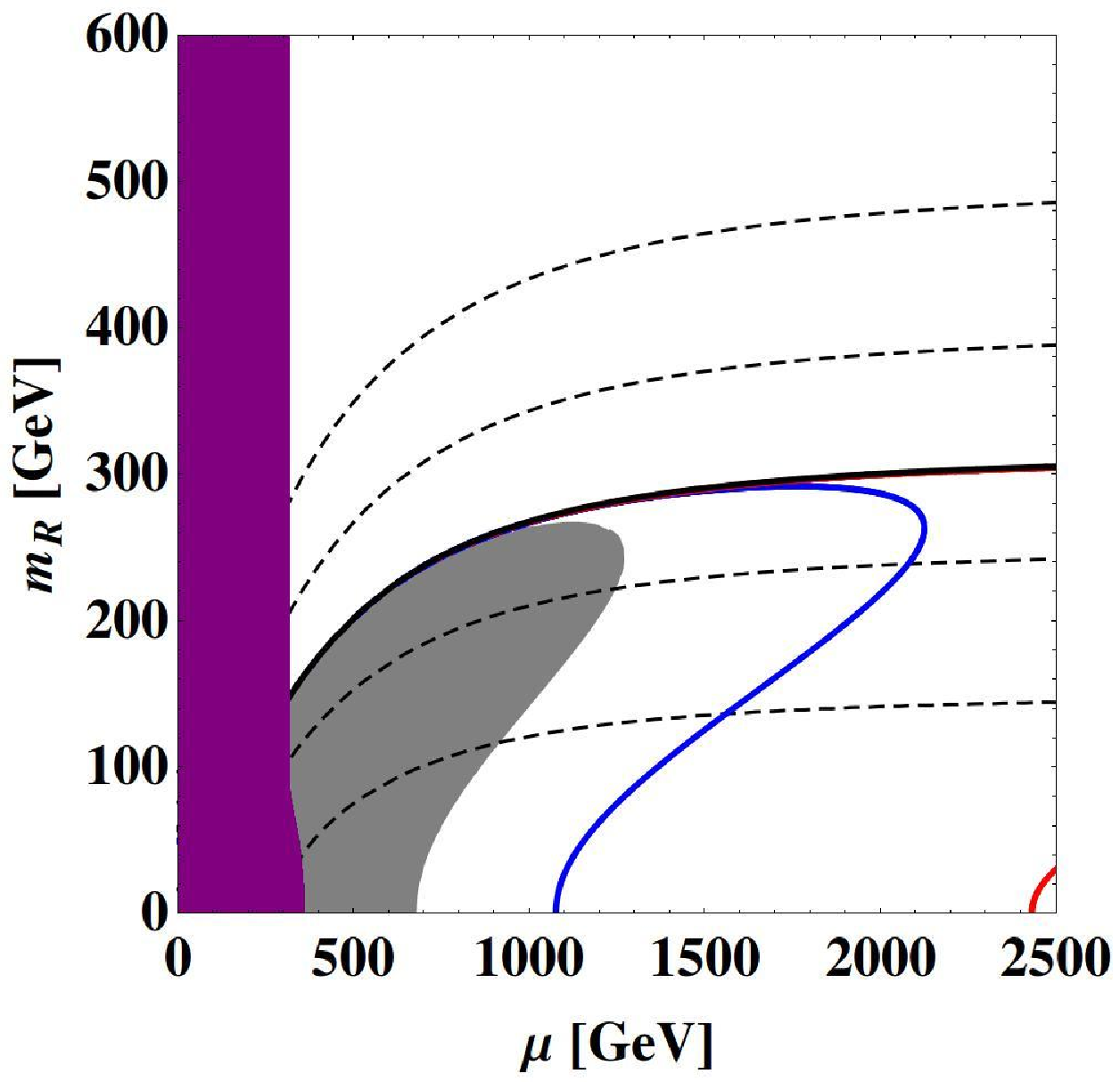}
  \end{minipage} 
\end{tabular}
 \caption{
The same as in Fig.~\ref{fig:Higgs:inv:decay},
but with the adjoint Yukawa couplings
$\lambdaad=0.4$.
%
	}
\label{fig:higgs:inv:adjoint} 
\end{figure}

The reason for this behavior is that
lighter $\phi$ implies smaller $g_{h\phi\phi}$ coupling.
As a comparison,
we also show in Fig.~\ref{fig:higgs:inv:adjoint}
the corresponding bound in the case of $\lambdaad=0.4$. 
In this case, the would-be pseudo-moduli direction 
is deformed by adjoint couplings
and consequently the constraints from the invisible decay
become slightly tight.
We see that a larger parameter region can be probed in a future,
especially when the soft mass $m_R$ is small.

As a reference,
we show two sets of sample parameters
in Table~\ref{tab:input:output}.
We take the averaged Higgsino mass $\mu$ equal to $\MS$,
while the ratio is given by $\mu_u/\mu_d=\tan\varphi/\tan\beta=1.25$.
The values of $\tan\alpha$ correspond to our input that 
the pseudo-scalar Higgs mass is equal to $\MS$.
The value of $\tan\alpha$ follows from taking 
the pseudo-scalar Higgs mass equal to $\MS$,
and implies $0.2\%$ enhancement ($0.8\%$ reduction)
of the up-type (down-type) Yukawa interactions.
The adjoint scalar masses according to 
$m^2_{\sigma_a}/8=8m^2_{\pi_a}=m^2_{D_a}$ for each $a=1,2,3$
lead to the $D$-term suppression $\suppress{D}=1/2$.

\begin{table}[htbp]
\begin{center}
\begin{tabular}{c|cccccc}
\hline
	\makebox[12mm]{Input} & 
	\makebox[18mm]{	$\MS\left(=\mu\right)$} & 
	\makebox[10mm]{$\tan\beta$} &	
	\makebox[22mm]{$M_{D_{1,2,3}\!}\units{TeV}$} & 
	\makebox[22mm]{$m_{R_{0,u,,d}\!}\units{GeV}$} & 
	\makebox[14mm]{	$ \tan\varphi $} &
	\makebox[12mm]{	$ \tan \alpha $} 
	\\\hline 
	Case 1 &
	  $1\nunits{TeV}$  &  $2$ &  $(1.0,\,3.0,\,7.0) $  & 
	$(24,\,120,\,120)$ &  $2.5$ &  $-0.45$ 
	\\
	Case 2 &
        $2\nunits{TeV}$ &  	$2$ & $(2.0,\,3.0,\,4.8) $  & 
	$(\ 0,\ 260,\,260) $ & $2.5$  & $-0.46$ \\
	\hline \hline 
	Output &
	$\Lambda\units{GeV}$ & 
	$\lambda (\MS)$ &
	$\suppress{D}$ & 
	$\suppress{R}$ & 
	$m_\phi $ &
        $\mathrm{Br}_{h\to\mathrm{inv}}$
        \\ \hline 
 Case 1 & $1.0\times 10^{17}$  
        & $0.73$ & $0.5$ & $0.02$ & $28\nunits{GeV}$ & $11\%$ 
 \\ 
 Case 2 & $4.0\times 10^{16}$ 
        & $0.69$ & $0.5$ & $0.02$ & $15\nunits{GeV}$ & $13\%$ 
 \\\hline
\end{tabular}
\end{center}
\caption{
	Sample sets of input/output parameters. 
The suppression factors $\suppress{D}$, $\suppress{R}$ 
and the mixing angle $\tan\varphi$ are defined 
by Eqs.~\eq{eq:Dterm:pot}, \eq{eq:suppressionR}
and \eq{eq:angle:def}, respectively.
}
\label{tab:input:output}
\end{table}


\section{Summary and Discussion}
\label{sec:summary}

We have studied a singlet extension of
the minimal $R$-symmetric SUSY SM with Dirac gauginos
that is consistent with gauge coupling unification.
Specifically we have calculated the mass of the SM-like Higgs boson
and found that 
its observed value of $125\nunits{GeV}$ can well be reproduced
in a small $\tan\beta$ region,
by the nMSSM-like, singlet Yukawa coupling $\lambda$
within perturbativity up to the unification scale
$10^{16}$--$10^{17}\nunits{GeV}$.
This is true even when the SUSY scale is as low as $1\nunits{TeV}$
and, remarkably, even without the standard $D$-term contribution
to the quartic Higgs potential.
The latter is important because
the $D$-term potential is known to be suppressed
in theories of Dirac gauginos and supersoft SUSY breaking.

The unification is preserved in a minimal way
by adding two vector-like pairs of singlet ``leptons''.
Adding more extra matters makes the gauge couplings stronger at UV
and thus relaxes the triviality bound.
We also examined the precision of unification by using two-loop RGE's.
Although we have not taken into account
the full variety of SUSY particle threshold,
we found that
a heavier Dirac gluino and/or wino makes the unification precise.
We note that
the mass threshold of a Dirac gaugino combined with
its scalar partners, $\sigma$ and $\pi$,
can conveniently be represented at one-loop by a single mass scale 
$M_D=\left(m_D^8 m_\sigma m_\pi\right)^{1/10}$.

The allowed parameter space is rather limited
in the small $\tan\beta$ region
especially when $SU(2)\times U(1)$ $D$-terms are suppressed
by adjoint scalar contributions.
Recent dedicated calculations 
\cite{Goodsell:2014pla,Diessner:2015yna}
show, however, that a part of two-loop contributions
give substantial improvement to the Higgs mass
in the singlet extensions of the MSSM;
This is also the case in models that contains Dirac gluino.
It would imply that
our calculation based on RG method
may underestimate the upper bound of the Higgs mass
and that there is broader parameter region
that is consistent with the triviality bound.

A characteristic feature of the present $R$-symmetric model
is the existence of light scalar and fermion modes, 
pseudo-moduli $\phi$ and pseudo-Goldstino $\psi$,
whose properties are restricted by the approximate $U(1)_R$ symmetry.
Although the mixing between the pseudo-moduli and the SM Higgs
is negligible,
it does affect the RG evolution of the SM Higgs quartic coupling
below the SUSY scale.
We have examined this effect 
by constructing
the SM coupled with the pseudo-moduli
and found that
the Higgs mass is reduced by a few \%,
but there exists a parameter region
consistent with $125\nunits{GeV}$ Higgs,
again even without the $D$-term.

As we discussed,
the constraints from invisible decays of the $Z$ and Higgs bosons
are weak if the singlet component dominates the lightest mass eigenstates.
The constraint from the $Z$ decay can easily be satisfied
since the present model contains the Higgsino mass parameters.
On the other hand,
the invisible Higgs decay generically
gives a tight constraint, 
as was originally expected.
We have found, however, that
there are interesting regions of parameter space
in which the Higgs invisible width is within the current bound
and within the proposed reaches of the future experiments:
This occurs especially when 
the Higgsino mass parameters are much larger
than $\lambda v$ of the order of $100\nunits{GeV}$.

There remain many issues to be discussed.
First, improved calculations 
including the full SUSY spectrum 
and higher-loop corrections are desired.
The invisible Higgs decay 
in the $R$-symmetric setup
should be studied in more general manner;
we have explored a limited region of parameter space.
For instance, if the singlet soft mass is larger than the doublet ones,
the lightest mass eigenstate $\phi$
deviates from the original pseudo-moduli direction.
This implies that 
the $\phi$ becomes heavier but has a larger coupling to the Higgs.
Loop corrections may also be important
since the suppression of the pseudo-moduli coupling
to the Higgs boson depends on a specific structure
of the potential.
Detailed studies on the adjoint Yukawa couplings
and their upper bounds should be done.
On the theoretical side,
it is also important to justify the assumptions
about the origin or the absence of superpotential terms
that we mentioned in \S\ref{sec:setup}.
Finally we should remark that
the cosmology of the singlet extended $R$-symmetric model
is still challenging and deserves further study.


\section*{Acknowledgment}

\noindent
The authors would like to thank 
Takashi~Shimomura, Osamu~Seto and Toshifumi~Yamashita
for discussions and useful comments on 
visible SUSY breaking, cosmological constraints and grand unification.
They also thank Teppei~Kitahara and Yutaro~Shoji
for discussions on the singlet extensions.
They are also grateful to
Takuya~Kakuda and Masahiro~Soeta
for collaboration at an early stage of the present work.
H.N. was supported in part by the Grant-in-Aid for the Ministry of 
Education, Culture, Sports, Science, and Technology of Japan 
No.~26400243.


\appendix


\section{Notes on RG Analysis}
\label{sec:RGanalysis}

Here we summarize necessary tools for RG analysis:
input parameters, one- and two-loop RGE's
below and above the SUSY scale $\MS$.
Two-loop RGE's are generated  by the Mathematica package 
SARAH~\cite{Staub:2008uz,Staub:2012pb,Staub:2013tta}. 
We write two-loop beta function
for a generic coupling $g_i$ in the form
$dg_i/dt=\beta^{(1)}_{g_i}/\left(16\pi^2\right)
+\beta^{(2)}_{g_i}/\left(16\pi^2\right)^2$,
where $t=\log\left(Q/m_Z\right)$ 
is a logarithm of the renormalization scale $Q$.
We mainly use $g_1^2=\left(5/3\right)g_Y^2$ for $U(1)$ coupling.

\subsection{Input Parameters}
\label{sec:inputs}

The input parameters at the EW scale are~\cite{Agashe:2014kda}
\begin{align} 
	\alpha_s(m_Z) &= 0.1184\nunits{GeV}\ ,\\
	m_h &= 125.7\nunits{GeV}\ ,\\ 
	m_t &= 173.21 \pm 0.87\nunits{GeV} \ . 
\end{align}
We use the center value for $m_t$
because its error gives negligibly small effects in our analysis.
The parameters at the top mass scale in the $\wb{\mathrm{MS}}$ scheme 
are given by~\cite{Buttazzo:2013uya}  
\begin{align} 
 g_s(m_t) 
        &=1.1666  
        - 0.000\,46 \left(\frac{m_t}{\mathrm{GeV}}-173.10\right) 
        \ ,\\
 g_2(m_t)
 &= 0.648\,22 + 0.000\,04 \left(\frac{m_t}{\mathrm{GeV}}-173.10\right)
 \, \\
 g_Y(m_t)
 &= 0.357\,61 + 0.000\,11 \left(\frac{m_t}{\mathrm{GeV}}-173.10\right) 
 \ ,\\
 y_t(m_t) 
        &= 0.935\,58 + 0.005\,50\left(\frac{m_t}{\mathrm{GeV}}-173.10 \right)
            \ ,\\
\frac{1}{2} \lambda_H(m_t) 
        &= 0.127\,11 
        + 0.002\,06\left(\frac{m_h}{\mathrm{GeV}}-125.66\right)
        -0.000\,04 \left(\frac{m_t}{\mathrm{GeV}} - 173.10\right)
        \ .
\end{align}



\subsection{SUSY RGE's with Singlet and Adjoint Yukawa Couplings}
\label{sec:RGE:ad:1-loop}

Here we show the relevant RGE's in our $R$-symmetric SUSY model that 
include the singlet and adjoint Yukawa couplings,
$\lambda$ and $\lambda_{ai}$,
defined by the superpotentials 
\eq{eq:W:Higgs} and \eq{eq:W:adjoint}, respectively.
We neglect the bottom and tau Yukawa couplings
since eventually we are interested in a small $\tan\beta$ region.
Then one-loop beta functions 
for the top Yukawa coupling $y_t$
and the singlet Yukawa coupling $\lambda$ are given by
\begin{align} 
\beta ^{(1)}_{y_t}
 =&\ y_t 
    \left({}
       6y_t^2
      + \lambda^2
      +|\lambda ^u_S|^2 +3 |\lambda ^u_T|^2
      -\frac{16}{3}g_3^2 -3 g_2^2 -\frac{13}{15}g_1^2
    \right)
    \ ,
\\
\beta ^{(1)}_{\lambda }
 =&\ \lambda 
    \left({}
       3y_t^2 
      + 4\lambda^2
      + |\lambda ^u_S|^2 +  |\lambda ^d_S|^2 
      +3|\lambda ^u_T|^2 + 3|\lambda ^d_T|^2
      -3 g_2^2 -\frac{3}{5}g_1^2
    \right)
    \ ,
\end{align}
and for the adjoint Yukawa couplings 
$\lambda_{ai}$ ($a=S,T$ and $i=u,d$),
\begin{align} 
\beta ^{(1)}_{\lambda^u_T }
 =&\ \lambda^u_T 
    \Bigl({}
       \lambda^2 
      + 2  |\lambda^u_S|^2 + 2 |\lambda^d_T|^2 + 8 |\lambda^u_T|^2
      - \frac{3}{5}g_1^2 - 7 g_2^2  
      + 3 y_t^2
    \Bigr)
    \ ,
\\
\beta ^{(1)}_{\lambda^d_T }
 =&\ \lambda^d_T 
    \Bigl({}
        \lambda^2
       +2  |\lambda^{d}_S|^2 +8 |\lambda^d_T|^2  +2  |\lambda^{u}_T|^2 
       - \frac{3}{5}g_1^2  -7 g_2^2  
    \Bigr)
    \ ,
\\
\beta ^{(1)}_{\lambda^u_S }
 =&\ \lambda^u_S 
    \Bigl({}
       \lambda^2 
      + 2 |\lambda^d_S|^2 +4 |\lambda^u_S|^2 +6 |\lambda^u_T|^2
      - \frac{3}{5}g_1^2  -3 g_2^2 
	 + 3 y_t^2
    \Bigr)
    \ ,
\\
\beta ^{(1)}_{\lambda^d_S }
 =&\ \lambda^d_S 
    \Bigl({}
       \lambda^2 
      + 4 |\lambda^d_S|^2 +2 |\lambda^u_S|^2 +6 |\lambda^d_T|^2  
      - \frac{3}{5}g_1^2  -3 g_2^2  
    \Bigr)
    \ .
\end{align}
%
%
%
%
%
%
We write our two-loop beta functions in the form
$\beta^{(2)}_{g_i}  =\beta ^{(2)}_{g_i, {\rm MDGSSM}} + \Delta \beta ^{(2)}_{g_i}$,
where $\beta ^{(2)}_{g_i, {\rm MDGSSM}}$ are beta functions 
in the minimal Dirac Gaugino SUSY SM 
(MDGSSM) \cite{Benakli:2014cia} 
with $U(1)_R$-violating couplings there,
$\lambda _S$, $\lambda _T$, $\lambda _{SR}$, 
$Y_{\hat{\tilde{E}}i}$, $Y^{ij}_{{\rm LFV}}$, $Y^j_{{\rm EFV}}$, 
$\lambda _{ST}$, $\lambda _{SO}$, $\kappa _o$,
as well as extra lepton couplings
$\lambda _{S\hat{E}ij}$ and $Y_{\hat{E}i}$,
all set to zero.
The deviations $\Delta \beta ^{(2)}_{g_i}$ are found as follows:
For gauge couplings $g_a$ ($a=1,2,3$), 
\begin{align}
\Delta \beta^{(2)}_{g_1}
\ ={}-\frac{6}{5}\lambda ^2 g^3_1
      \ , \qquad
\Delta \beta^{(2)}_{g_2}
\ ={}-2 \lambda^2 g^3_2 
      \ , \qquad
\Delta \beta^{(2)}_{g_3}
\ =\ 0 \ ,
\end{align}
and for the MSSM Yukawa couplings,
\begin{align}
\Delta \beta ^{(2)}_{Y_u}
 & = \lambda Y_u 
     \Bigl[ 
     - 3\lambda ^2 -3 |\lambda^d_T|^2 - |\lambda^d_S|^2
     - Y^\dagger _d Y_d
     - 3 Y_{u}^{\dagger}  Y_u
     - 3\Tr{Y_d Y_d^{\dagger}}
     -  \Tr{Y_e Y_e^{\dagger}}
     \Bigr] \ ,
\\
\Delta \beta ^{(2)}_{Y_d}
 & = \lambda Y_d 
     \Bigl[
     - 3\lambda ^2  -3 |\lambda^u_T|^2 - |\lambda^u_S|^2
     -   {Y_u^{\dagger} Y_u}  
     - 3 {Y_d^{\dagger} Y_d}
     - 3 \Tr{Y_u Y_u^{\dagger}}
     \Bigr] \ ,
 \\
 \Delta \beta^{(2)}_{Y_e}
 & = \lambda Y_e 
     \Bigl[
     - 3\lambda ^2 -3 |\lambda^u_T|^2 - |\lambda^u_S|^2
     - 3 {Y_e^{\dagger} Y_e}
     - 3\Tr{Y_u Y_u^{\dagger}}
     \Bigr]
     \ .
\end{align}
For the adjoint Yukawa couplings $\lambda_{ai}$, we have
\begin{align}
\Delta 	\beta_{\lambda^u_T}^{(2)} 
 & = 
     \lambda ^2 \lambda^u_T 
     \Bigl[
     - 3 \lambda ^2 -5 |\lambda^d_T|^2 -5 |\lambda^u_T|^2 
     - |\lambda^d_S|^2 - |\lambda^u_S|^2 
     - 3 \Tr{Y_d Y_d^{\dagger}}
     - \Tr{Y_e Y_e^{\dagger}}
     \Bigr] \ ,
\\
\Delta 	\beta_{\lambda^d_T}^{(2)}
 & = 
     \lambda ^2 \lambda^d_T 
     \Bigl[
     - 3 \lambda ^2 -5 |\lambda^d_T|^2 -5 |\lambda^u_T|^2 
     - |\lambda^d_S|^2 - |\lambda^u_S|^2 
     - 3 \Tr{Y_u Y_{u}^{\dagger}}
     \Bigr] \ ,
\\
\Delta 	\beta_{\lambda^u_S}^{(2)} 
 & = 
     \lambda ^2 \lambda^u_S
     \Bigl[
     -3 \lambda ^2 -3 |\lambda^d_T|^2 -3 |\lambda^u_T|^2 
     - 3|\lambda^d_S|^2 - 3|\lambda^u_S|^2 
     -3 \Tr{Y_d Y_d^{\dagger}}
     -  \Tr{Y_e Y_e^{\dagger}}
     \Bigr]
     ,
\\
\Delta 	\beta_{\lambda^d_S}^{(2)}
 & = 
     \lambda ^2 \lambda^d_S 
     \Bigl[
     - 3 \lambda ^2 -3 |\lambda^d_T|^2 -3 |\lambda^u_T|^2 
     - 3|\lambda^d_S|^2 - 3|\lambda^u_S|^2 
     -3 \Tr{Y_u  Y_{u}^{\dagger}}
     \Bigr] \ .
\end{align}
Finally and most importantly, 
for the singlet Yukawa coupling $\lambda$,
we have
\begin{align}
 \beta_{\lambda}^{(2)} 
\ =&\  
     \beta_{\lambda}^{(2,0)} 
    +\beta_{\lambda}^{(2,1)} 
    +\beta_{\lambda}^{(2,2)} 
     \ ,
\\
 \beta_{\lambda}^{(2,0)} 
  =&{}
     - 10 \lambda^{5} 
     +2 \left(
       \frac{3}{5} g_{1}^{2} 
       + 3 g_{2}^{2}
      \right) \lambda^{3}
     + \frac{3}{2}
       \left(
         \frac{99}{25} g_{1}^{4} 
       + \frac{6}{5} g_{1}^{2} g_{2}^{2}
       + 11 g_{2}^{4} 
        \right) \lambda 
      \ ,
\end{align}
where $\beta_{\lambda}^{(2,1)}$ and $\beta_{\lambda}^{(2,1)}$ 
are contributions from the adjoint and the MSSM Yukawa couplings,
\begin{align}
 \beta_{\lambda}^{(2,1)} 
 =&{}
     - 3\lambda^3
       \Bigl(
         3 |\lambda^u_T|^2
        +3 |\lambda^d_T|^2
        +  |\lambda^u_S|^2 
        +  |\lambda^d_S|^2 
       \Bigr)
     - 3 \lambda 
       \Bigl(
         5 |\lambda^u_T|^4 
       + 5 |\lambda^d_T|^4 
       +   |\lambda^u_S|^4 
       +   |\lambda^d_S|^4 
       \Bigr)
\nonumber\\
  &{}
     - 4\lambda 
     \Bigl(
        3 |\lambda^u_T|^2 |\lambda^d_T|^2  
      +   |\lambda^u_S|^2 |\lambda^d_S|^2 
      \Bigr)
     - 6\lambda 
     \Bigl(
        |\lambda^u_S|^2 |\lambda^u_T|^2
      + |\lambda^d_S|^2 |\lambda^d_T|^2 
      \Bigr)
\nonumber\\
  &{}
     + 12 \lambda g_2^2 
       \Bigl(
            |\lambda^u_T|^2 
          + |\lambda^d_T|^2
       \Bigr)
      \ ,
\\
 \beta_{\lambda}^{(2,2)} 
 =&{}
    - 3\lambda^3
      \left[
          3 \Tr{Y_u Y_u^{\dagger}}
         +3 \Tr{Y_d Y_{d}^{\dagger}}
         +  \Tr{Y_e Y_{e}^{\dagger}}
       \right]
\nonumber \\ 
  &  - 3\lambda 
       \left[
           3 \Tr{Y_u Y_u^{\dagger}  Y_u Y_u^{\dagger}}
         + 3 \Tr{Y_d Y_d^{\dagger}  Y_d Y_d^{\dagger}}
         + 2 \Tr{Y_d Y_u^{\dagger}  Y_u Y_d^{\dagger}}
         +   \Tr{Y_e Y_e^{\dagger}  Y_e Y_e^{\dagger}}
       \right]
\nonumber \\ 
  &{}
    + 16 \lambda g_3^2 
      \left[
            \Tr{Y_u Y_u^{\dagger}}
          + \Tr{Y_d Y_d^{\dagger}}
      \right]
\nonumber \\ 
  &{} 
    + \frac{2}{5} \lambda g_1^2 
      \left[
            \Tr{Y_d Y_d^{\dagger}}
        + 2 \Tr{Y_u Y_u^{\dagger}}
        + 3 \Tr{Y_e Y_e^{\dagger}}
      \right]
    \ .
\end{align}

\subsection{
RGE's in the SM coupled with a complex scalar}
\label{sec:RGE:CSSM}


Here we show two-loop RGE's
in the SM coupled with a complex scalar $\phi$,
whose quartic potential is given by \Eq{eq:CSSM}.
In our case,
such theory is used in \S{\ref{subsec:match2pmoduli}}
as a low-energy effective theory of our $R$-symmetric model
containing the light pseudo-moduli $\phi$.

For the effective theory couplings 
$g_i=\lambda_H$, $\lambda_{\phi H}$ and $\lambda_\phi$,
one-loop beta functions $\beta ^{(1)}_{g_i}$ are shown 
in \Eqs{eq:RGE:lambdaH:oneloop}{eq:RGE:ytSM:oneloop}.
For the SM couplings,
we write two-loop beta functions 
in the form
$\beta^{(2)}_{g_i} =\beta ^{(2)}_{g_i, {\rm SM}} + \Delta \beta ^{(2)}_{g_i}$,
where the SM contribution 
can be found for instance 
in Refs.~\cite{Luo:2002ey,Buttazzo:2013uya}.
For the gauge couplings $g_a$ ($a=3,2,1$),
the Yukawa coupling matrices $Y_f$ ($f=u,d,e$)
and the Higgs quartic coupling $\lambda_H$,
the new contributions $\Delta \beta ^{(2)}_{g_i}$ are found to be
\begin{align}
\Delta \beta_{g_a}^{(2)}
\ =\ 0
     \ , \qquad
\Delta \beta_{Y_f}^{(2)}
\ =\ \frac{1}{2} Y_f \lambda_{\phi H}^{2}
     \ , \qquad
\Delta \beta ^{(2)}_{\lambda_H}
\ ={}-10 \lambda_H \lambda_{\phi H}^{2} -8 \lambda_{\phi H}^{3}
      \ .
\end{align}
For the couplings $\lambda_{\phi H}$ and $\lambda_\phi$
in the effective theory \eq{eq:CSSM},
we have 
\begin{align}
\beta ^{(2)}_{\lambda_{\phi H}} 
 =&   \left(
        \frac{1671}{400} g_1^4 
       +\frac{9}{8} g_1^2 g_2^2 
       -\frac{145}{16} g_2^4 
      \right)
      \lambda_{\phi H} 
     +\left(
        \frac{3}{5} g_1^2  + 3 g_2^2 
      \right)
      \Bigl(
         12 \lambda_H 
        +   \lambda_{\phi H} 
      \Bigr)
      \lambda_{\phi H}
\nonumber \\
 &{}
     -\lambda_{\phi H} 
      \left[
        15 \lambda_H^2 
        +\frac{5}{2} \lambda_{\phi}^2
        +12\Bigl(
           3 \lambda_H + \lambda_{\phi} 
           \Bigr)
        \lambda_{\phi H}
        +11 \lambda_{\phi H}^2
      \right]
\nonumber\\
 &{}+ \lambda_{\phi H} \Tr{Y_u Y_u^{\dagger}}
      \left[
          \frac{17}{4} g_1^2 
        + \frac{45}{4} g_2^2
        + 40 g_{3}^{2}
        -12\Bigl(
             3 \lambda_H 
             +\lambda_{\phi H} 
           \Bigr) 
      \right] 
\nonumber \\ 
 &{}+\lambda_{\phi H} \Tr{Y_d Y_d^{\dagger}} 
     \left[
        \ \frac{5}{4} g_1^2 + \frac{45}{4} g_2^2 + 40 g_3^2 
       -12\Bigl(
             3 \lambda_H 
             +\lambda_{\phi H} 
          \Bigr) 
     \right]
\nonumber \\
 &{}+\lambda_{\phi H} \Tr{Y_e Y_e^{\dagger}}
     \left[
       \frac{15}{4} g_1^2  + \frac{15}{4} g_2^2  
      -4 \Bigl(
           3 \lambda_H  + \lambda_{\phi H}
          \Bigr)
     \right] 
     - 21 \lambda_{\phi H} \Tr{Y_d Y_u^{\dagger} Y_u Y_d^{\dagger}}
\nonumber\\
 &{}-\frac{9}{2} \lambda_{\phi H} 
     \left[
       3\Tr{Y_u Y_u^{\dagger} Y_u Y_u^{\dagger}}
      +3\Tr{Y_d Y_d^{\dagger} Y_d Y_d^{\dagger}}
      + \Tr{Y_e Y_e^{\dagger} Y_e Y_e^{\dagger}}
    \right]
       \ , 
\\
\beta_{\lambda_{\phi}}^{(2)} 
  =&{}- \Bigl(
           15 \lambda_{\phi}^3
         + 20 \lambda_{\phi} \lambda_{\phi H}^2
         + 32 \lambda_{\phi H}^3 
        \Bigr)
      + 16\lambda_{\phi H}^2 
        \left( 
            \frac{3}{5} g_1^2
          + 3 g_2^2 
        \right) 
\nonumber\\
 &  {}- 16 \lambda_{\phi H}^{2} 
        \left[
            3 \Tr{Y_u Y_u^{\dagger}}
          + 3 \Tr{Y_d Y_d^{\dagger}}
          +   \Tr{Y_e Y_e^{\dagger}}
        \right]
        \ .
\end{align}



\section{A Note on Dirac Gaugino Mass Threshold}
\label{sec:note:gaugino}


Let us consider an $SU(N)$ gauge theory that contains
a gaugino $\lambda_a$, its Dirac partner $\chi_a$,
and their scalar partners, $A_a=\left(\sigma_a+i\pi_a\right)/\sqrt{2}$,
all in the adjoint representations.
We are interested in the mass threshold
in the one-loop running of the gauge coupling constant.
We will show that
the mass threshold can be represented 
by a single mass threshold $M_D$ defined by
\begin{align}
M_D
\ =\ \left[m_D^8 m_\sigma m_\pi\right]^{1/10}
     \ .
\label{eq:MD:def}
\end{align}
To see this, 
let us look at the solution to one-loop gauge RGE
in the form
\begin{align}
\frac{2\pi}{\alpha\fun{Q}}-\frac{2\pi}{\alpha\fun{Q_0}}
\ =\ b_0\ln\frac{Q_0}{Q}
    +\sum_i \Delta_i b\,\ln\frac{m_i}{Q}
     \ ,
\end{align}
where $b_0$ is the massless contribution 
to the beta function coefficient,
and $\Delta_i b$ is a contribution
of the $i$-th particle of mass $m_i$.
The sum is taken over all the particles
whose mass lies between the renormalization scale $Q$
and the reference scale $Q_0$ at IR.
For the case of interest,
a Dirac pair of gaugino $\lambda_\alpha$ and $\chi_\alpha$
contributes $\Delta_D b=4N/3$ 
while its scalar partners give
$\Delta_\sigma b=\Delta_\pi b=N/6$.
If these fields were degenerate in mass,
$m_\sigma=m_\pi=m_D$,
we would have
\begin{align}
\sum_{i=\lambda,\chi,\sigma,\pi}
 \Delta_i b\ln\frac{m_i}{Q}
\ =\ \left(\frac{4N}{3}+\frac{2N}{6}\right)\ln\frac{m_D}{Q}
\ =\ \frac{5N}{3}\,\ln\frac{m_D}{Q}
     \ .
\end{align}
Actually we have
\begin{align}
\sum_{i=\lambda,\chi,\sigma,\pi}
 \Delta_i b\ln\frac{m_i}{Q}
\ =\ \frac{4N}{3}\,\ln\frac{m_D}{Q}
    +\frac{N}{6}\,\ln\frac{m_\sigma}{Q}
    +\frac{N}{6}\,\frac{m_\pi}{Q}
\ =\ \frac{5N}{3}\,\ln\frac{M_D}{Q}
     \ .
\end{align}
We see that at the leading log level,
the mass threshold effect
due to the massive adjoint fields in a Dirac gaugino model
can be represented
by a single mass threshold scale $M_D$
defined by \Eq{eq:MD:def}.

\def\suppression{\mathcal{Z}}

If we denote the holomorphic and non-holomorphic
contributions to the adjoint scalar mass term
by $b$ and $m^2_A$, respectively,
the squared masses
of the real and imaginary components of the adjoint scalars are
\begin{align}
m^2_\sigma
 &\equiv\  \suppression_\sigma m_D^2
\ =\ 4m_D^2 +b+m^2_A
    \ ,
\nonumber\\
m^2_\pi
 &\equiv\ \suppression_\pi m_D^2
\ ={}- b+m^2_A
    \ .
\end{align}
In principle these adjoint scalar masses can 
take any values,
(although it involves a fine tuning).
Therefore we can regard the coefficients $\suppression_{\sigma,\pi}$
as (positive) free parameters.
Using the above parametrization,
we have a relation
$M_D=\left(\suppression_\sigma\suppression_\pi\right)^{1/10}m_D$.

The Dirac mass threshold $M_D$ defined here
coincides with the actual Dirac mass of the gaugino
if $\suppression_\sigma\suppression_\pi=1$, that is,
\begin{align}
m^2_\sigma
\ =\ \suppression_\sigma m^2_D 
     \ , \qquad
m^2_\pi
\ =\ \frac{1}{\suppression_\sigma}\,m^2_D 
     \ ,
\end{align}
Otherwise,
the mass threshold scale does not coincide with
the Dirac mass parameter itself.
Notice that
requiring that $m_D=m_\sigma=m_\pi$ would imply
a negative mass correction to the real scalar $\sigma$,
$\delta m^2_\sigma=m^2_A+b=-3m_D^2$.


\section{Examples of Gauge Coupling Unification}
\label{sec:unification}


Here we give several examples in which 
gauge coupling unification is 
satisfied
under our simplifying assumption
stated in \S{\ref{subsec:unification}}.
We use two-loop RGE's summarized in App.~\ref{sec:RGanalysis}
and
 treat the SUSY threshold
by a single representative scale $\MS$
with the exception
of the extra vector-like leptons and 
$SU(3)\times SU(2)$ Dirac gauginos.
The results are summarized in Table~\ref{tab:gauge}.

\begin{table}[bp]
\begin{center}
\begin{tabular}{c||c|ccc}
	Case &
	\makebox[10mm]{	$\MS $} & 
	\makebox[14mm]{$\ME\units{TeV}$ } & 
	\makebox[26mm]{$M_{D_{1,2,3}\!}\units{TeV}$} & 
	$\Lambda\units{GeV}$  
	\\\hline\hline 
	(I)&
	$1\nunits{TeV}$  &
	$10$--$15$ &  $(1.0,\,1.0,\,1.0) $  & $1.0\times 10^{17}$ 
	\\	
	(II)&
	&  $1$ &  $(1.0,\,1.0,\,3.0) $  & $3.0\times 10^{16}$ 
	\\ \hline	
	(III)&
	$2\nunits{TeV}$ 
	&  	$2$ & $(1.0,\,1.0,\,6.0) $  & $2.0\times 10^{16}$
	\\
	(IV)&
	&  	$6$ & $(2.0,\,3.0,\,4.8) $  & $4.0\times 10^{16}$
	\\
\hline
(V)&
	$1\nunits{TeV}$ 
	&  $6$ &  $(1.0,\,1.5,\,2.4) $  & $1.0\times 10^{17}$ 
	\\	
	(VI)&
	&  $5$ &  $(1.0,\,3.0,\,7.0) $  & $5.0\times 10^{16}$ 
 \\
 \hline
\end{tabular}
\end{center}
\caption{
	Sample sets of parameters for gauge coupling unification;
$\suppressionD{D}=0$, $\tan\beta=2$.
}
\label{tab:gauge}
\end{table}

\begin{figure}[tbp]
 \begin{tabular}{lr}
  \begin{minipage}{0.45\hsize}
   \begin{center}
	\includegraphics[scale=0.8]{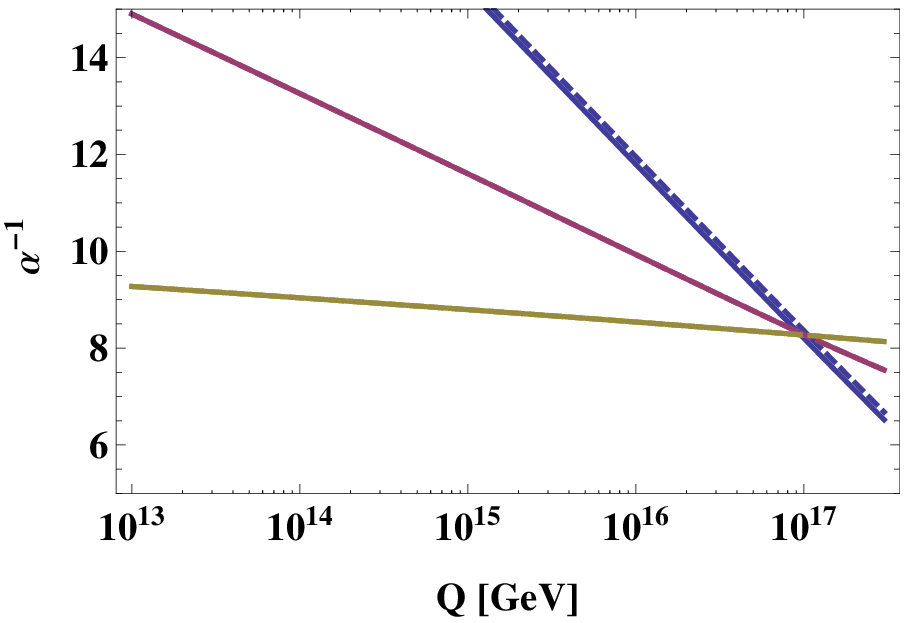}
   \end{center}
  \end{minipage}
  \hspace{4mm}
  \begin{minipage}{0.45\hsize}
   \begin{center}
	\includegraphics[scale=0.8]{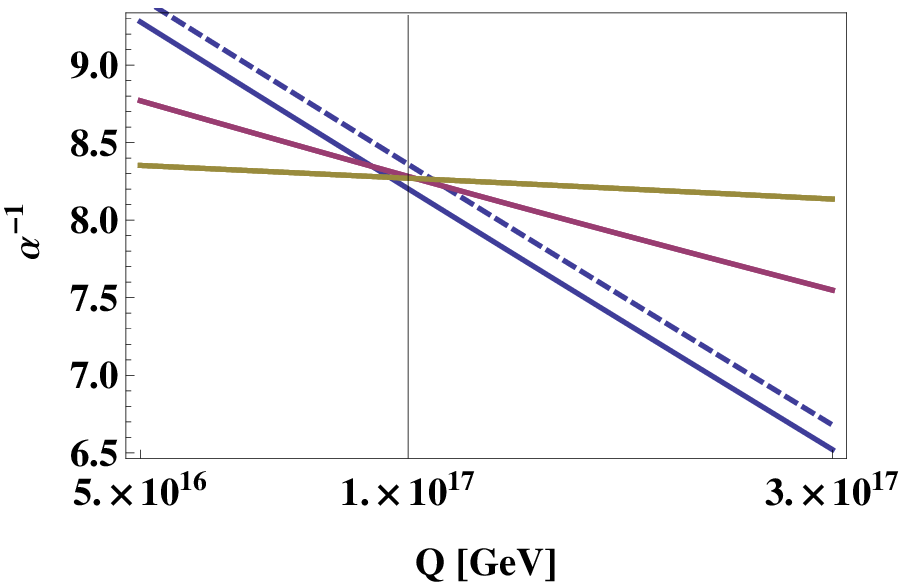}
   \end{center}
  \end{minipage}
 \end{tabular}
\caption{
Two-loop running of gauge couplings
corresponding to parameter set (I) in Table~\ref{tab:gauge}.
The extra lepton mass is 
$\ME=10\nunits{TeV}$ (solid) and $15\nunits{TeV}$ (dashed).
}
\label{fig:g-run:2-loop:hel:fin}
\end{figure}

In the first example,
the unification is achieved
by tuning the bachelor mass $\ME$.
Recall that two-loop contributions make
the $SU(3)$ gauge coupling slightly asymptotically non-free.
This implies that
the unification scale $\Lambda$ becomes slightly larger than
the one-loop value $2\times 10^{16}\nunits{GeV}$.
With the extra leptons heavier,
$U(1)_Y$ gauge coupling becomes smaller
so that
it can cross the intersecting point of $SU(3)\times SU(2)$ couplings.
This corresponds to parameter set (I) in Table~\ref{tab:gauge} 
and is depicted in Fig.~\ref{fig:g-run:2-loop:hel:fin}.

\begin{figure}[tbp]
 \begin{tabular}{lr}
  \begin{minipage}{0.45\hsize}
   \begin{center}
	\includegraphics[scale=0.8]{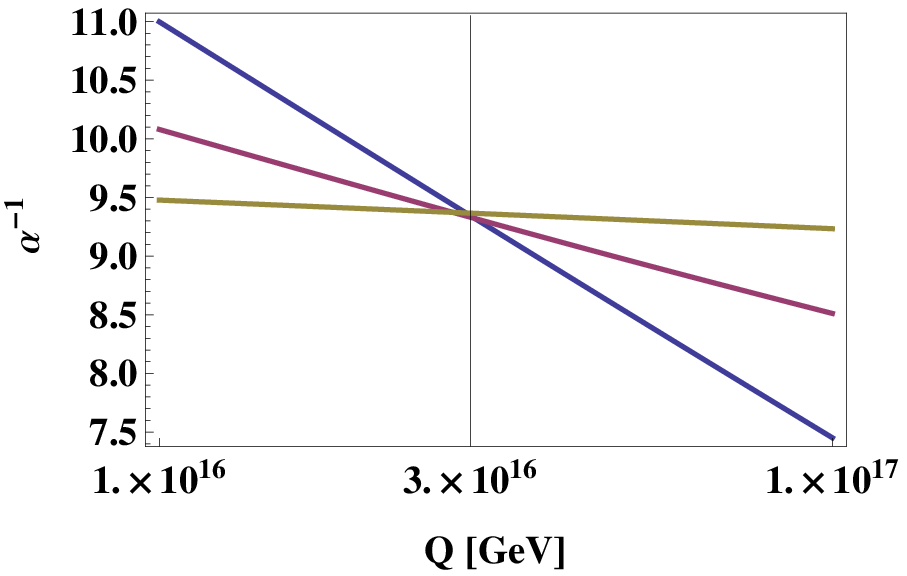}
\caption{
Two-loop gauge running 
corresponding to parameter set (II):
$M_{D_3}=3\nunits{TeV}$ and $M_{D_2}=\MS=1\nunits{TeV}$.
}
\label{fig:g-run:2-loop:hgluino:fin}
   \end{center}
  \end{minipage}
  \hspace{8mm}
  \begin{minipage}{0.45\hsize}
   \begin{center}
        \includegraphics[scale=0.8]{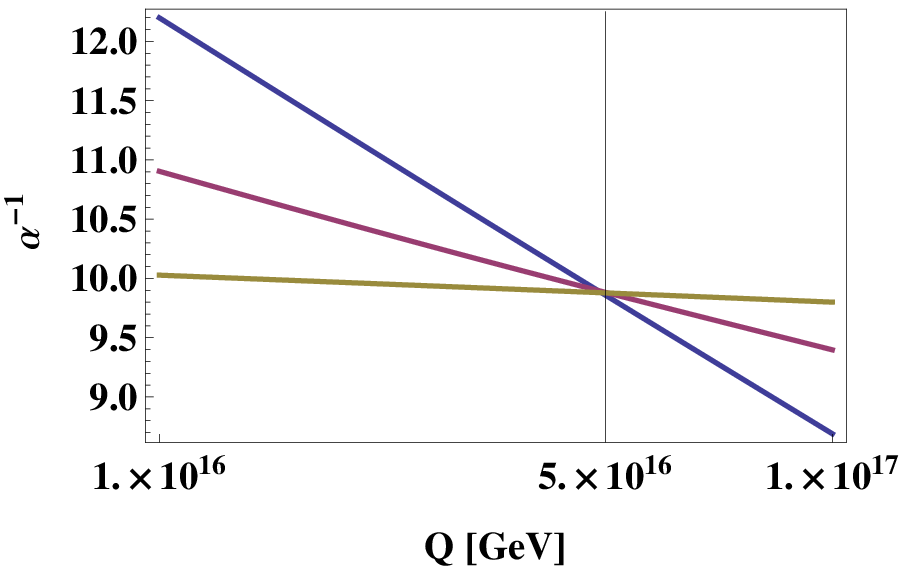}
 \caption{
Two-loop gauge running 
corresponding to parameter set (VI):
$M_{D_3}:M_{D_2}:M_{D_1}=7:3:1\nunits{TeV}$.
}
\label{fig:g-run:2-loop:gaugino}
  \end{center}
  \end{minipage}
 \end{tabular}
\end{figure}

More natural examples are provided
by changing the Dirac mass thresholds, $M_{D_3}$ and $M_{D_2}$, 
for the $SU(3)\times SU(2)$ gauginos.  
Although $M_{D_a}$ is not the same as the gaugino mass $m_{D_a}$, 
it is plausible that
Dirac gluino is the heaviest and Dirac Wino is the next heaviest gaugino,
due to an RG effect.
A heavier Dirac gluino mass threshold
implies a smaller $SU(3)$ coupling at UV.
By tuning the extra lepton mass,
unification can be achieved
at relatively lower energy scale around $10^{16}\nunits{GeV}$.
In this way we obtain
parameter sets (II) and (III) in Table~\ref{tab:gauge}.
Two-loop running gauge couplings in parameter set (II)
are shown in Fig.~\ref{fig:g-run:2-loop:hgluino:fin}.

In passing, it is interesting to note that
under our simplifying assumption,
the extra leptons must be lighter than $\MS=1\nunits{TeV}$
if $M_{D3}$ is larger than $3\nunits{TeV}$;
similarly,
the extra leptons must be lighter than $\MS=2\nunits{TeV}$
if $M_{D3}>6\nunits{TeV}$.

Finally we consider the case in which
the $SU(2)$ gaugino mass threshold can also be different:
parameter sets (IV), (V) and (VI) in Table~\ref{tab:gauge}.
In this case it is not so easy to discuss the effect qualitatively
because changing $M_{D_2}$ affects the running of $SU(3)$ coupling.
Parameter set (IV) is shown 
in Fig.~\ref{fig:g-run:2-loop:gaugino}.

%



  
\end{document}